\documentclass[12pt]{article}
\usepackage{amsmath}
\usepackage{graphicx}
\usepackage{enumerate}
\usepackage{natbib}
\usepackage{amsfonts}
\usepackage{mathrsfs}
\usepackage{dsfont}
\usepackage{bbding}
\usepackage{amssymb}
\numberwithin{equation}{section}

\DeclareMathOperator{\Soft}{Soft}
\usepackage{amsmath}
\usepackage{setspace}
\newtheorem{theorem}{Theorem}
\newtheorem{lemma}{Lemma}

\newtheorem{remark}{Remark}
\newtheorem{condition}{Condition}
\usepackage{graphicx}
\usepackage{booktabs}
\usepackage{color}
\usepackage{multirow}
\usepackage{subfigure}
\usepackage{algorithm}
\usepackage{algorithmic}
\usepackage[colorlinks,
           allcolors=blue,
           ]{hyperref}
\newcommand{\blind}{1}

\allowdisplaybreaks[1]
\begin{document}

\def\spacingset#1{\renewcommand{\baselinestretch}%
{#1}\small\normalsize} \spacingset{1}

\if1\blind
{
  \title{\bf A General Framework of Online Updating Variable Selection for Generalized Linear Models  with Streaming Datasets}
  \author{Xiaoyu Ma,  Lu Lin \thanks{
    The author gratefully acknowledges \textit{the National Natural Science Foundation of China (Grant No. 11971265) of China.}} \footnote{The corresponding
author. Email: lulinsdu@gmail.com}\hspace{.2cm}
    \\
   \small{ Zhongtai Securities Institute for Financial Studies, Shandong University}\\
     and \\
    Yujie Gai\thanks{
    The author gratefully acknowledges \textit{the National Natural Science Foundation of China (Grant No. 12071497) and National Statistical Science Research Project (Grant No. 2020LY014).}}\hspace{.2cm}\\
   \small{ School of Statistics and Mathematics, Central University of Finance and Economics}
    }
   \date{}
  \maketitle
} \fi

\if0\blind
{
  \bigskip
  \bigskip
  \bigskip
  \begin{center}
    {\LARGE\bf A General Framework of Online Updating Variable Selection for Generalized Linear Models  with Streaming Datasets}
\end{center}
  \medskip
} \fi

\bigskip
\begin{abstract}
In the research field of big data, one of important issues is how to recover the sequentially changing sets of true features when the data sets arrive sequentially. The paper presents a general framework for online updating variable selection and parameter estimation in generalized linear models with streaming datasets. This is a type of online updating penalty likelihoods with differentiable or non-differentiable penalty function. The online updating coordinate descent algorithm is proposed to solve the online updating optimization problem. Moreover, a tuning parameter selection is suggested in an online updating way. The selection and estimation consistencies, and the oracle property are established, theoretically. Our methods are further examined and illustrated by various numerical examples from both simulation experiments and a real data analysis.
\end{abstract}
\noindent%
{\it Keywords:} Streaming data environment; Penalty likelihoods; Online updating BIC criterion; Oracle property. 
\vfill
\newpage
\spacingset{1.5} 
\section{Introduction}
\label{sec:intro}
A data set that arrives in streams and chunks  is termed a streaming data set. In the era of big data, the streaming data sets from various areas such as bioinformatics, medical imaging and computer vision are rapidly increasing in volume and velocity. This brings us challenges to learn efficient statistical models and inferences. How to make statistical inferences without storage requirement for previous raw data is the key in the streaming data environment.

Assume we have the data set $\left\{(\boldsymbol{x}_{ti},y_{ti}), i=1,\cdots, n_t; t=1,\cdots,b\right\}$ up to the $b$-th batch, where $n_t$ is the the sample size at the $t$-batch and the total sample size is $N_b=\sum_{t=1}^{b}n_t$. Denote by $D_t=\{\boldsymbol{y}_t,\boldsymbol{X}_t \}$ the $t$-th batch data set, where $\boldsymbol{y}_{t}=(y_{t1},\cdots,y_{tn_t})^{T}$, $\boldsymbol{X}_t=(\boldsymbol{x}_{t1},\cdots,\boldsymbol{x}_{tn_t})^{T}$. We suppose that $(y_{ti},\boldsymbol{x}_{ti})$ for all $t$ and $i$ are independent and identically distributed observations of $(y, \boldsymbol{X})$ with response $y \in \mathrm{R}$ and covariate $\boldsymbol{X} \in \mathrm{R}^{p}$. To our knowledge, amounts of methodologies have been developed for related parameter estimation and statistical inference for the case of low-dimensional covariate $\boldsymbol{X}$, especially for the case when the dimension $p$ is less than every batch sample sizes $n_t, t \in \{1,\cdots,b\}$. \cite{R1951} proposed the well-known stochastic gradient descent (SGD) algorithm that has been extensively used in the filed of machine learning. It is obvious that the SGD algorithm is applied to the streaming data because of using a part of gradients in every steps. However, the SGD method is not robust to learning rate, and it may fail to converge if the learning rate is too large. As improvements, the implicit SGD (ISGD) algorithm and averaged implicit SGD (AISGD) algorithm were developed by \cite{T2014}, which could be used in the streaming data to estimate model parameters. Another type of recursive algorithm, the cumulative updating method has been well developed, including the online least squares estimator (OLSE) for linear model by \cite{S1994}, the cumulative estimating equation (CEE) estimator and the cumulatively updating estimating equation (CUEE) estimator of \cite{S2016} for non-linear models. \cite{L2002} proposed an updating weighted sum for two samples. Recently, \cite{L2020b} proposed a renewable estimation and inference for generalized linear model, and \cite{L2020a} introduced a general framework of renewable weighted sums for various online updating estimation.

Although the methods aforementioned attain favourable estimation and achieve satisfactory theoretical conclusions, they need the condition that the dimension $p$ of covariate is smaller than every batch sample sizes $n_t, t \in \{1,\cdots,b\}$. For the case of high-dimensional covariate, an important issue is how to select important variables in an online updating framework. Several algorithms were developed for the variable selection in the models with streaming data sets. \cite{X2010} developed regularized dual averaging (RDA) method to solve the general online regularized problem, which can select active variables via appropriate penalty functions. \cite{L2009}proposed a variant of the truncated SGD.  \cite{F2018} applied the truncated SGD to linear model and gave the statistical analysis of the truncated SGD. \cite{S2020} introduced a novel framework for variable selection in linear model, which combines the updated statistics and truncation techniques. These SGD algorithms and truncation techniques, however, are sensitive to learning rate or step size, and tend to select the set with larger cardinality for including all important variables.
 Moreover, the related works are applied  mainly to linear models, without the well-developed methodology for general models.

In this paper, we suggest a renewable optimized objective function for parameter estimation and variable selection in generalized linear models (GLMs) with high-dimensional covariates. The new method is an online updating strategy for general penalty likelihoods, including smooth penalty likelihoods and non-smooth penalty likelihoods. This optimized objective function with various penalty functions could select important variables efficiently. In order to realize numerical solution, we introduce efficient algorithms for this optimization problem. Our method enjoys a fast convergence rate and can efficiently recover the support of the true signals. Moreover, the proposed method can choose tuning parameters via a data-driven and online updating BIC criterion. Theoretically, the method achieves the estimation and selection consistencies and oracle property. It can be seen from the theoretical conclusions that the method is free of the constraint on the number of batches, which means that the new method is adaptive to the situation where streaming data sets arrive fast and perpetually.

The remainder of this paper is organized as follows. Section 2 presents our renewable optimized objective function and concludes the key large sample estimation and selection properties. Section 3 proposes incremental updating algorithm to compute renewable selection and estimation results. Section 4 reports main simulation studies and the real data analysis. Concluding remarks are provided in Section 5. Proofs of the theorems and all technical details are relegated to Supplementary Material.
\section{Methodologies and Main Results}
\label{sec:methods}
\subsection{Setup}
Suppose we have the streaming data sets $\left\{(\boldsymbol{x}_{ti},y_{ti}), i=1,\cdots,n_t; t=1,\cdots,b\right\}$ satisfying the GLM: $E(y_{ti}|\boldsymbol{x}_{ti})=g(\boldsymbol{x}_{ti}^{T}\boldsymbol{\beta})$, where $g(\cdot)$ is a known link function, $\boldsymbol{\beta} \in \mathrm{R}^{p}$ is the unknown parameter vector, and all the data $(\boldsymbol{x}_{ti},y_{ti})$ are independent and identically distributed from the distribution \begin{equation}\label{density} f(y;\boldsymbol{X}, \boldsymbol{\beta}_{0}, \phi_0)=\exp\left\{\frac{y\boldsymbol{X}^{T}\boldsymbol{\beta}_0-
b(\boldsymbol{X}^{T}\boldsymbol{\beta}_0)}{a(\phi_0)}+c(y,\phi_0)\right\},\end{equation} where $\boldsymbol{\beta}_0 \in  \mathrm{R}^{p}$ is the true value of the parameter of interest and $\phi_0$ is the true value of a nuisance parameter. We permit the dimension of covariate $\boldsymbol{X}$ is large than the sample sizes $n_t$, and there is no a pre-decided size relationship between total sample size $N_b$ and the dimension $p$ in the streaming data sets. We suppose that only a part of components of covariate $\boldsymbol{X}$ affect the response. Denote by $S=\{j \in \{1,\cdots,p\}  | \beta_j  \ne 0 \}$ the index set of active covariates and by $s=|S|$ the cardinality of the set $S$. Our goal is then to select the active index set $S$ and estimate the corresponding parameters $\beta_j, j \in S$ under the environment of streaming data.

\subsection{Renewable Optimized Objective Function}
 Denote by $D_{b}^{\star}=\left\{D_{1}, \ldots, D_{b}\right\}$ the cumulative data set up to the $b$-batch. For the GLM, we write the associated log-likelihood function as:
 \begin{align*}
 \ell\left(\boldsymbol{\beta}, \phi ; D_{b}^{\star}\right)=\sum_{t=1}^{b}\sum_{i=1}^{n_t} \log f\left(y_{ti} ; \boldsymbol{x}_{ti}, \boldsymbol{\beta}, \phi\right)
 =\sum_{t=1}^{b}\sum_{i=1}^{n_t}\frac{y_{ti}\boldsymbol{x}_{ti}^{T}\boldsymbol{\beta}-b(\boldsymbol{x}_{ti}^{T}\boldsymbol{\beta})}{a(\phi)}+c(y_{ti},\phi).
\end{align*}
It is equivalent to $\sum_{t=1}^{b}\sum_{i=1}^{n_t}\left\{y_{ti}\boldsymbol{x}_{ti}^{T}\boldsymbol{\beta}-b(\boldsymbol{x}_{ti}^{T}\boldsymbol{\beta})\right\}$ when we maximize $\ell\left(\boldsymbol{\beta}, \phi ; D_{b}^{\star}\right)$ with respect to $\boldsymbol{\beta}$.
For convenience, slightly abusing the notation, we use $\ell_{t}\left(\boldsymbol{\beta}; D_{t}\right):=\left\{\boldsymbol{y}_{t}^{T}\right.\boldsymbol{X}_{t}\boldsymbol{\beta}-
\left.\boldsymbol{1}^{T}b(\boldsymbol{X}_{t}\boldsymbol{\beta})\right\}$ to represent  the simplified log-likelihood function in the $t$-th batch for any $t \in \{1, \cdots, b\}$. Denote the score function as $\boldsymbol{U}_t(\boldsymbol{\beta}; D_t):=\nabla_{\boldsymbol{\beta}}\ell_{t}(\boldsymbol{\beta}; D_t)$, and the negative Hessian matrix as $\boldsymbol{J}_t(\boldsymbol{\beta}; D_t):=-\nabla_{\boldsymbol{\beta}}\boldsymbol{U}_t(\boldsymbol{\beta}; D_t)$ for any $t \in \{1,\cdots,b\}$, where $\nabla_{\boldsymbol{\beta}}$ stands for the derivative of a function with respect to $\boldsymbol{\beta}$.

Under high dimension and sparsity situation, by commonly used techniques, we establish a penalty likelihood by adding a penalty term into the likelihood function. That is to say, we estimate the parameters by solving the following penalty likelihood:
\begin{equation*}
\begin{split}
\hat{\boldsymbol{\beta}}_b=\arg\min_{\boldsymbol{\beta} \in \mathrm{R}^{p}}\left\{\sum_{t=1}^{b} \left\{-\ell_{t}\left(\boldsymbol{\beta}; D_{t}\right)+P(\boldsymbol{\beta};\lambda_t)\right\}\right\},
\end{split}
\end{equation*}
where $P(\boldsymbol{\beta};\lambda_{t})$ is a known penalty function. Here the penalty function $P(\boldsymbol{\beta};\lambda_{t})$ can be chosen to be differentiable or non-differentiable with respect to $\boldsymbol{\beta}$. We then denote by $\partial P(\boldsymbol{\beta};\lambda_t)$ the derivative of $P(\boldsymbol{\beta};\lambda_t)$ with respect to $\boldsymbol{\beta}$ if $P(\boldsymbol{\beta};\lambda_t)$ is differentiable, and use the same notation $\partial P(\boldsymbol{\beta};\lambda_t)$ to denote the subgradient if $P(\boldsymbol{\beta};\lambda_t)$ is non-differentiable.

In the case of streaming data sets, our first task is to represent the above penalty likelihood by an online updating form.
Assume we have got a consistent penalty estimator $\hat{\boldsymbol{\beta}}_{b-1}$ until the $(b-1)$-th batch. Then, by the Taylor expansions of $\ell_{t}\left(\boldsymbol{\beta}; D_{t}\right)$ and $P(\boldsymbol{\beta};\lambda_{t})$ for $t=1,\cdots,b-1$  around the  accumulative estimator $\hat{\boldsymbol{\beta}}_{b-1}$, we have
\begin{align*}
\hat{\boldsymbol{\beta}}_b=\arg&\min_{\boldsymbol{\beta} \in \mathrm{R}^{p}}\sum_{t=1}^{b-1}\left\{-\left[\ell_{t}(\hat{\boldsymbol{\beta}}_{b-1}; D_{t})+(\boldsymbol{\beta}-\hat{\boldsymbol{\beta}}_{b-1})^{T}\boldsymbol{U}_t(\hat{\boldsymbol{\beta}}_{b-1}; D_{t})\right.\right.\\
&\left.-\frac{1}{2}(\boldsymbol{\beta}-\hat{\boldsymbol{\beta}}_{b-1})^{T}\boldsymbol{J}_t(\hat{\boldsymbol{\beta}}_{b-1}; D_{t})(\boldsymbol{\beta}-\hat{\boldsymbol{\beta}}_{b-1})+
o_p(\|\boldsymbol{\beta}-\hat{\boldsymbol{\beta}}_{b-1}\|)
\right]\\
&+\left.\left[P(\hat{\boldsymbol{\beta}}_{b-1};\lambda_t)+(\boldsymbol{\beta}-\hat{\boldsymbol{\beta}}_{b-1})^{T}
\partial P(\hat{\boldsymbol{\beta}}_{b-1};\lambda_t)+
o_p(\|\boldsymbol{\beta}-\hat{\boldsymbol{\beta}}_{b-1}\|)\right]\right\}\\&+\left\{-\ell_{b}\left(\boldsymbol{\beta}; D_{b}\right)+P(\boldsymbol{\beta};\lambda_b)\right\}\\
&=\arg\min_{\boldsymbol{\beta} \in \mathrm{R}^{p}} \left\{ \sum_{t=1}^{b-1}\left\{-\ell_{t}(\hat{\boldsymbol{\beta}}_{b-1}; D_{t})+P(\hat{\boldsymbol{\beta}}_{b-1};\lambda_t)\right\}\right.\\
&+(\boldsymbol{\beta}-\hat{\boldsymbol{\beta}}_{b-1})^{T}\sum_{t=1}^{b-1}\left\{-\boldsymbol{U}_t(\hat{\boldsymbol{\beta}}_{b-1}; D_{t})+\partial P(\hat{\boldsymbol{\beta}}_{b-1};\lambda_t)\right\}\\
&+\frac{1}{2}(\boldsymbol{\beta}-\hat{\boldsymbol{\beta}}_{b-1})^{T}\sum_{t=1}^{b-1}\boldsymbol{J}_t(\hat{\boldsymbol{\beta}}_{b-1}; D_{t})(\boldsymbol{\beta}-\hat{\boldsymbol{\beta}}_{b-1})\\
&\left.-\ell_{b}\left(\boldsymbol{\beta}; D_{b}\right)+P(\boldsymbol{\beta};\lambda_b)+o_p(\|\boldsymbol{\beta}-\hat{\boldsymbol{\beta}}_{b-1}\|)
\right\}.
\end{align*} Under some mild regularity conditions, when $n_{b-1}$ is large enough, the error term $o_p(\|\boldsymbol{\beta}-\hat{\boldsymbol{\beta}}_{b-1}\|)$ can be ignored. Note that the accumulative estimator $\hat{\boldsymbol{\beta}}_{b-1}$ satisfies $\sum_{t=1}^{b-1}\{-\boldsymbol{U}_{t}(\hat{\boldsymbol{\beta}}_{b-1},D_t)+\partial P(\hat{\boldsymbol{\beta}}_{b-1};\lambda_t)\}=0$. Ignoring these constant terms and error term $o_p(\|\boldsymbol{\beta}-\hat{\boldsymbol{\beta}}_{b-1}\|)$, the above optimization problem can be expressed as following form:
\begin{equation*}
\begin{split}
\hat{\boldsymbol{\beta}}_b&=\arg\min_{\boldsymbol{\beta} \in \mathrm{R}^{p}}\frac{1}{2}(\boldsymbol{\beta}-\hat{\boldsymbol{\beta}}_{b-1})^{T}\left\{\sum_{t=1}^{b-1}\boldsymbol{J}_t(\hat{\boldsymbol{\beta}}_{b-1}; D_{t})\right\}(\boldsymbol{\beta}-\hat{\boldsymbol{\beta}}_{b-1})-\ell_{b}\left(\boldsymbol{\beta}; D_{b}\right)+P(\boldsymbol{\beta};\lambda_b).
\end{split}
\end{equation*}
If each sample size $n_t$ is large enough, we have $\hat{\boldsymbol{\beta}}_{b-1}\approx\hat{\boldsymbol{\beta}}_{t}$.
Replacing $\hat{\boldsymbol{\beta}}_{b-1}$ with $\hat{\boldsymbol{\beta}}_{t}$ in $\boldsymbol{J}_t(\hat{\boldsymbol{\beta}}_{b-1}; D_{t})$ leads to the following incremental updating expression:
\begin{equation}\label{obj_opt}
\begin{split}
\hat{\boldsymbol{\beta}}_b=\arg\min_{\boldsymbol{\beta} \in \mathrm{R}^{p}}\frac{1}{2}(\boldsymbol{\beta}-\hat{\boldsymbol{\beta}}_{b-1})^{T}\left\{\sum_{t=1}^{b-1}\boldsymbol{J}_t(\hat{\boldsymbol{\beta}}_{t};D_{t})\right\}&(\boldsymbol{\beta}-\hat{\boldsymbol{\beta}}_{b-1})\\
&-\ell_{b}\left(\boldsymbol{\beta}; D_{b}\right)+P(\boldsymbol{\beta};\lambda_b),
\end{split}
\end{equation}
Obviously, the objective function in \eqref{obj_opt} has online updating structure, which only uses the summary statistics $\boldsymbol{J}_t(\hat{\boldsymbol{\beta}}_{t};D_{t})$, $\hat{\boldsymbol{\beta}}_{b-1}$ and the raw data in the last batch. Thus, it can be used for variable selection and parameter estimation in the model with streaming data sets.

\subsection{Theoretical Properties}

In this subsection, we establish the estimation and selection consistencies, and the oracle properties for the proposed methods. Suppose that $\{(y_{ti},\boldsymbol{x}_{ti}), i=1,\cdots,n_t; t=1,\cdots,b\}$ are i.i.d. samples from the exponential dispersion model (\ref{obj_opt}).
Write $\boldsymbol{\beta}_0=(\beta_{01},\cdots,\beta_{0p})^{T}=(\boldsymbol{\beta}_{01}^{T},
\boldsymbol{\beta}_{02}^{T})^{T}$ with $\boldsymbol{\beta}_{01} \in \mathrm{R}^{s}$ and $\boldsymbol{\beta}_{02} \in \mathrm{R}^{p-s}$. Without loss of generality, assume that $\boldsymbol{\beta}_{02}=\boldsymbol{0}$. Let $I(\boldsymbol{\beta}_{0})$ be the true Fisher information matrix and $I_1(\boldsymbol{\beta}_{01},\boldsymbol{0})$ be the true Fisher information matrix knowing $\boldsymbol{\beta}_{02}=\boldsymbol{0}$. Assume the penalty function has the additive form: $P(\boldsymbol{\beta};\lambda_b)=\sum_{j=1}^{p}p(\beta_j;\lambda_b)$. This additivity condition is mild and most classical penalty functions have this property, see for example the penalty functions in LASSO (\cite{T1996}), SCAD (\cite{F2001}) and MCP(\cite{Z2010}).
Throughout this section, we write $p_{\lambda_{N_b}}(\cdot)$ for $p(\cdot;\lambda_b) $ to emphasize $\lambda_b$ depending on the sample size $N_b$. Denote $L_b(\boldsymbol{\beta})=\frac{1}{2}(\boldsymbol{\beta}-\hat{\boldsymbol{\beta}}_{b-1})^{T}\left\{\sum_{t=1}^{b-1}\boldsymbol{J}_t(\hat{\boldsymbol{\beta}}_{t};D_{t})\right\}(\boldsymbol{\beta}-\hat{\boldsymbol{\beta}}_{b-1})-\ell_{b}\left(\boldsymbol{\beta}; D_{b}\right)$, $Q_b(\boldsymbol{\beta})=-L_b(\boldsymbol{\beta})-N_b\sum_{j=1}^{p}p_{\lambda_{N_b}}(\beta_j)$,
and $\hat{\boldsymbol{\beta}}_b$ is the  local maximizer of $Q_b(\boldsymbol{\beta})$.


We postulate the following regularity conditions.

\begin{condition}
\label{con1}
 The Fisher information matrix $I_{N_b}(\boldsymbol{\beta})=\frac{1}{N_b}\sum_{t=1}^{b}\sum_{i=1}^{n_t}b^{''}(\boldsymbol{x}_{ti}^{T}\boldsymbol{\beta})\boldsymbol{x}_{ti}\boldsymbol{x}_{ti}^{T}/a(\phi)$ is finite and positive definite for all $\boldsymbol{\beta}$ in some neighborhood of $\boldsymbol{\beta}_0$.
\end{condition}

\begin{condition}
\label{con2}
 For all $\boldsymbol{\beta}$ in some neighborhood of $\boldsymbol{\beta}_0$, there is $|b^{(3)}(\boldsymbol{X}^{T}\boldsymbol{\beta})| \leq M_0(\boldsymbol{X})$ for some functions $M_0(\boldsymbol{X})$ satisfying $E\{M_0(\boldsymbol{X})X_jX_kX_l\} < \infty$ for all $j,k,l$.
\end{condition}

\begin{condition}
\label{con3}
The log-likelihood function $\ell(\boldsymbol{\beta}, \phi, \boldsymbol{X}; y)$ is twice continuously differentiable and $I_{N_b}(\boldsymbol{\beta})$ is Lipschitz continuous in parameter space $\Theta$.
\end{condition}

\begin{remark}
The condition \ref{con1} and condition \ref{con2} are the standard regularity conditions, see, for example, \cite{L1983}. The condition \ref{con3} is needed for renewable estimators, see \cite{L2020b}.
\end{remark}

We then have the following estimation consistency.

\begin{theorem}[Estimation Consistency]
\label{the1}
Under conditions \ref{con1}-\ref{con3}, if $\max\{|p_{\lambda_{N_b}}^{''}(|\beta_{0j}|)|: \beta_{0j} \neq 0 \} \to 0$, then $\hat{\boldsymbol{\beta}}_b$  satisfies $\|\hat{\boldsymbol{\beta}}_b-\boldsymbol{\beta}_0\|=O_P(N_{b}^{-1/2}+a_{N_b})$, where $a_{N_b}=\max\{p_{\lambda_{N_b}}^{'}(|\beta_{0j}|): \beta_{0j} \neq 0 \}$.
\end{theorem}

It is clear from Theorem \ref{the1} that by choosing a proper $\lambda_{N_b}$, the online updating estimator is root-$N_b$ consistent. The proof of Theorem \ref{the1} is given in Supplementary Material. We now show that this estimator possesses the sparsity property $\hat{\boldsymbol{\beta}}_{b2}=0$ and the oracle property. To this end, we need following lemma.

\begin{lemma}
\label{lem1}
Under conditions \ref{con1}-\ref{con3}, assume that
\begin{equation}
\label{con4}
\liminf_{n \to \infty}\liminf_{\theta \to 0^{+}} p_{\lambda_{N_b}}^{'}(\theta)/\lambda_{N_b} >0,
\end{equation}
where $\theta$ is the argument of the function. 
If $\lambda_{N_b} \to 0$ and $\sqrt{N_b}\lambda_{N_b} \to \infty$ as $N_b \to \infty$, then with probability tending to $1$, for any given $\boldsymbol{\beta}_{1}$ satisfying $\|\boldsymbol{\beta}_{1}-\boldsymbol{\beta}_{01}\|=O_P(N_b^{-1/2})$ and any constant $C$,
$$
Q_b\left\{\left(
\begin{aligned}
\boldsymbol{\beta}_{1} \\
\mathbf{0}
\end{aligned}
\right)
\right\}=\max_{\|\boldsymbol{\beta}_2\| \leq CN_b^{-1/2}}Q_b\left\{\left(
\begin{aligned}
\boldsymbol{\beta}_{1} \\
\boldsymbol{\beta}_{2}
\end{aligned}
\right)
\right\}.
$$
\end{lemma}

The proof of Lemma \ref{lem1} is given in Supplementary Material. This lemma also
implies the selection consistency.

\begin{theorem}[ Sparsity and Oracle Property]
\label{the2}
Under conditions \ref{con1}-\ref{con3},  assume that the penalty function $p_{\lambda_{N_b}}(|\theta|)$ satisfies condition \eqref{con4}. If $\lambda_{N_b} \to 0$ and $\sqrt{N_b}\lambda_{N_b} \to \infty$ as $N_b \to \infty$, then with probability tending to $1$, the root-$N_b$ consistent estimator $\hat{\boldsymbol{\beta}}_b=\left(\begin{aligned}&\hat{\boldsymbol{\beta}}_{b1}\\&\hat{\boldsymbol{\beta}}_{b2}
\end{aligned}
\right)$ in Theorem \ref{the1} satisfies

(a) Sparsity: $\hat{\boldsymbol{\beta}}_{b2}=\boldsymbol{0}$;

(b) Asymptotic normality:
$$
\begin{aligned}
\sqrt{N_b}\left\{\hat{\boldsymbol{\beta}}_{b1}-\boldsymbol{\beta}_{01}
\right\} \to N \left\{\boldsymbol{0},I_1^{-1}(\boldsymbol{\beta}_{01})
\right\}
\end{aligned}
$$
in distribution, where $I_1(\boldsymbol{\beta}_{01})=I(\boldsymbol{\beta}_{01},\boldsymbol{0})$, the Fisher information knowing $\boldsymbol{\beta}_{02}=\boldsymbol{0}$.
\end{theorem}

The proof of Theorem \ref{the2} is given in Supplementary Material. From the theorem, it can be seen that the sparsity property $\hat{\boldsymbol{\beta}}_{b2}=0$ holds, the online updating estimators for nonzero components have the asymptotic normality and the standard convergence rate of order $O_P(N_b^{-1/2})$, and the asymptotic covariance is the same as that of the estimator obtained by supposing that zero components $\boldsymbol{\beta}_{02}$ were known beforehand. Moreover, these theoretical results always hold without any constraint on $b$, the number of batches. It is an important property for the models with streaming data sets because it means that
the method is adaptive to the situation where streaming data sets arrive fast
and perpetually.

\section{Online Updating Coordinate Descent Algorithm}
\subsection{Online Updating Coordinate Descent}
For convenience, we denote the objective function in \eqref{obj_opt} as
$$L(\boldsymbol{\beta}; D_{b}^{\star} )=\frac{1}{2}(\boldsymbol{\beta}-\hat{\boldsymbol{\beta}}_{b-1})^{T}\left\{\sum_{t=1}^{b-1}\boldsymbol{J}_t(\hat{\boldsymbol{\beta}}_{t};D_{t})\right\}(\boldsymbol{\beta}-\hat{\boldsymbol{\beta}}_{b-1})
-\ell_{b}\left(\boldsymbol{\beta}; D_{b}\right)+P(\boldsymbol{\beta};\lambda_b).$$
The first derivative of  $L(\boldsymbol{\beta}; D_{b}^{\star} )$ with respect to $\boldsymbol{\beta}$ is
$$\frac{\partial L}{\partial \boldsymbol{\beta}}=\left\{\sum_{t=1}^{b-1}\boldsymbol{J}_t(\hat{\boldsymbol{\beta}}_{t};D_{t})\right\}\left(\boldsymbol{\beta}-\hat{\boldsymbol{\beta}}_{b-1}\right)-\boldsymbol{U}_b\left(\boldsymbol{\beta}; D_{b}\right)+ \partial P(\boldsymbol{\beta};\lambda_b).$$
By taking the first-order Taylor expansion of $\boldsymbol{U}_b\left(\boldsymbol{\beta}; D_{b}\right)$ around the  accumulative estimator $\hat{\boldsymbol{\beta}}_{b-1}$, and ignoring higher terms, we have
\begin{align*}
\frac{\partial L}{\partial \boldsymbol{\beta}}&\approx\left\{\sum_{t=1}^{b-1}\boldsymbol{J}_t(\hat{\boldsymbol{\beta}}_{t};D_{t})\right\}\left(\boldsymbol{\beta}-\hat{\boldsymbol{\beta}}_{b-1}\right)\\
&-\left\{\boldsymbol{U}_b\left(\hat{\boldsymbol{\beta}}_{b-1}; D_{b}\right)-
\boldsymbol{J}_b(\hat{\boldsymbol{\beta}}_{b-1};D_{b})\left(\boldsymbol{\beta}-\hat{\boldsymbol{\beta}}_{b-1}\right)\right\}+
\partial P(\boldsymbol{\beta};\lambda_b)\\
&=\left\{\sum_{t=1}^{b-1}\boldsymbol{J}_t(\hat{\boldsymbol{\beta}}_{t};D_{t})+\boldsymbol{J}_b(\hat{\boldsymbol{\beta}}_{b-1};D_{b})\right\}\left(\boldsymbol{\beta}-\hat{\boldsymbol{\beta}}_{b-1}\right)
-\boldsymbol{U}_b\left(\hat{\boldsymbol{\beta}}_{b-1}; D_{b}\right)+\partial P(\boldsymbol{\beta};\lambda_b)
\end{align*}
Let the first derivative $\frac{\partial L}{\partial \boldsymbol{\beta}}$ equal zero, we attain the following equation:
\begin{equation}\label{obj_equ}
\left\{\sum_{t=1}^{b-1}\boldsymbol{J}_t(\hat{\boldsymbol{\beta}}_{t};D_{t})+\boldsymbol{J}_b(\hat{\boldsymbol{\beta}}_{b-1};D_{b})\right\}\left(\boldsymbol{\beta}-\hat{\boldsymbol{\beta}}_{b-1}\right)
-\boldsymbol{U}_b\left(\hat{\boldsymbol{\beta}}_{b-1}; D_{b}\right)+\partial P(\boldsymbol{\beta};\lambda_b)
=0.
\end{equation} This is an online updating estimating equation.

Under the high dimensional setting, however, the Newton-Raphson algorithm is not practical because it is difficult or impossible to get the inverse of the negative Hessian matrix.
Motivated by \cite{F2010}, we solve the equation \eqref{obj_equ} via coordinate descent algorithm when the penalty function has the additive form: $P(\boldsymbol{\beta};\lambda_b)=\sum_{j=1}^{p}p(\beta_j;\lambda_b)$. As shown in the previous section, the additive penalty function is commonly used in the existing literature.
In the following, we take the LASSO penalty $P(\boldsymbol{\beta};\lambda_b)=\lambda_b\sum_{j=1}^{p}|\beta_j|, \lambda_b >0$ as an example to introduce the algorithm.
We begin with the scenario where $D_b$ arrives after $D_{b-1}$. In this case, there are the summary statistics $\sum_{t=1}^{b-1}\boldsymbol{J}_t(\hat{\boldsymbol{\beta}}_{t};D_{t})$, $\boldsymbol{J}_b(\hat{\boldsymbol{\beta}}_{b-1};D_{b})$, $\boldsymbol{U}_b(\hat{\boldsymbol{\beta}}_{b-1};D_{b})$ and the consistent estimator $\hat{\boldsymbol{\beta}}_{b-1}$. Consider a coordinate descent step from the $(b-1)$-th batch to the $b$-th batch for the equation for $\hat{\boldsymbol{\beta}}_b$:
\begin{equation}
\left\{\sum_{t=1}^{b-1}\boldsymbol{J}_t(\hat{\boldsymbol{\beta}}_{t};D_{t})+\boldsymbol{J}_b(\hat{\boldsymbol{\beta}}_{b-1};D_{b})\right\}\left(\hat{\boldsymbol{\beta}}_b-\hat{\boldsymbol{\beta}}_{b-1}\right)
-\boldsymbol{U}_b\left(\hat{\boldsymbol{\beta}}_{b-1}; D_{b}\right)+\lambda_b \partial \left\{\sum_{j=1}^{p}|\hat{\beta}_{b,j}|\right\}=0.
\end{equation}
Let estimators $\hat{\beta}_{b,l}=\hat{\beta}_{b-1,l}$ for $l \ne j$. Our goal is to solve the solution to $\hat{\beta}_{b,j}$. If $\hat{\beta}_{b,j} >0$, then
\begin{equation}
\left\{\sum_{t=1}^{b-1}\boldsymbol{J}_t(\hat{\boldsymbol{\beta}}_{t};D_{t})+\boldsymbol{J}_b(\hat{\boldsymbol{\beta}}_{b-1};D_{b})\right\}_{[j,j]}\left(\hat{\beta}_{b,j}-\hat{\beta}_{b-1,j}\right)
-\boldsymbol{U}_b\left(\hat{\boldsymbol{\beta}}_{b-1}; D_{b}\right)_{[j]}+\lambda_b \hat{\beta}_{b,j}=0,
\end{equation}
where the subscripts $[j,j]$ and $[j]$ represent the $j$-th row and $j$-th column element of the matrix and the $j$-th element of the vector, respectively.
We have the similar expression when $\hat{\beta}_{b,j} <0$, and $\hat{\beta}_{b,j} = 0$ is treated separately. Denote $Z_{b,j}=\boldsymbol{U}_b\left(\hat{\boldsymbol{\beta}}_{b-1}; D_{b}\right)_{[j]}+\hat{\beta}_{b-1,j}W_{b,j}$, and $W_{b,j}=\left\{\sum_{t=1}^{b-1}\boldsymbol{J}_t(\hat{\boldsymbol{\beta}}_{t};D_{t})+\boldsymbol{J}_b(\hat{\boldsymbol{\beta}}_{b-1};D_{b})\right\}_{[j,j]}$. Simple calculation shows that the coordinate-wise updating for LASSO has the form
\begin{equation}\label{value_LASSO}
\hat{\beta}_{b,j}^{LASSO}=\frac{\Soft(Z_{b,j}, \lambda_b)}{W_{b,j}},
\end{equation}
where $\Soft(z,\gamma)$ is the soft-thresholding operator with value
$$
\operatorname{sign}(z)(|z|-\gamma)_{+}=\left\{\begin{array}{ll}
z-\gamma & \text { if } z>0 \text { and } \gamma<|z|,\\
z+\gamma & \text { if } z<0 \text { and } \gamma<|z|, \\
0 & \text { if } \gamma \geq|z|.
\end{array}\right.
$$

We can get the similar expressions for SCAD and MCP penalties by the same way, which have the following forms respectively:
\begin{equation}\label{value_SCAD}
\left\{
\begin{aligned}
&W_{b,j}> \frac{1}{r-1}, \hat{\beta}_{b,j}^{SCAD}=\left\{
\begin{aligned}
&\frac{\Soft(Z_{b,j},\lambda_b)}{W_{b,j} }& &\text{ if } |Z_{b,j}| \leq \lambda_b+\lambda_bW_{b,j}\\
&\frac{\Soft(Z_{b,j},\frac{r\lambda_b}{r-1})}{(W_{b,j}-1/r)} & &\text{ if } \lambda_b+\lambda_bW_{b,j} < |Z_{b,j}|<r\lambda_bW_{b,j} \\
&Z_{b,j}/W_{b,j} & &\text{ if }  |Z_{b,j}| \geq r\lambda_bW_{b,j}
\end{aligned}\right. \\
&W_{b,j}< \frac{1}{r-1},  \hat{\beta}_{b,j}^{S
CAD}=\left\{
\begin{aligned}
&\frac{\Soft(Z_{b,j},\lambda_b)}{W_{b,j}} & &\text{ if } |Z_{b,j}| \leq r\lambda_bW_{b,j} \\
&\frac{\Soft(Z_{b,j},\frac{r\lambda_b}{r-1})}{(W_{b,j}-1/r)} & &\text{ if }   r\lambda_bW_{b,j} < |Z_{b,j}|<\lambda_b+\lambda_bW_{b,j} \\
&Z_{b,j}/W_{b,j} & &\text{ if }  |Z_{b,j}| \geq \lambda_b+\lambda_bW_{b,j}
\end{aligned}\right.\\
&W_{b,j}= \frac{1}{r-1},  \hat{\beta}_{b,j}^{SCAD}=\left\{
\begin{aligned}
&(r-1)\Soft(Z_{b,j},\lambda_b) & &\text{ if }  |Z_{b,j}| \leq \frac{r\lambda_b}{r-1}\\
&(r-1)Z_{b,j} & &\text{ if }   |Z_{b,j}| >\frac{r\lambda_b}{r-1},
\end{aligned}\right.
\end{aligned}\right.
\end{equation}

\begin{equation}\label{value_MCP}
\left\{
\begin{aligned}
&W_{b,j}> \frac{1}{r}, \hat{\beta}_{b,j}^{MCP}=\left\{
\begin{aligned}
&\frac{\Soft(Z_{b,j},\lambda_b)}{(W_{b,j}-1/r)} & &\text{ if } |Z_{b,j}| \leq r\lambda_bW_{b,j}\\
&Z_{b,j}/W_{b,j} & &\text{ if }  |Z_{b,j}| > r\lambda_bW_{b,j}
\end{aligned}\right. \\
&W_{b,j}< \frac{1}{r},  \hat{\beta}_{b,j}^{MCP}=\left\{
\begin{aligned}
&0 & &  \text{ if } |Z_{b,j}| \leq r\lambda_bW_{b,j}\\
&\frac{(Z_{b,j}-\lambda_b)}{(W_{b,j}-1/r) }& &\text{ if } r\lambda_bW_{b,j} < Z_{b,j} < \lambda_b \\
&\frac{(Z_{b,j}+\lambda_b)}{(W_{b,j}-1/r)} & &\text{ if } -\lambda_b < Z_{b,j} < -r\lambda_bW_{b,j} \\
&Z_{b,j}/W_{b,j} & &\text{ if }  |Z_{b,j}| \geq \lambda_b
\end{aligned}\right.\\
&W_{b,j}= \frac{1}{r},  \hat{\beta}_{b,j}^{MCP}=\left\{
\begin{aligned}
&0 & &\text{ if }  |Z_{b,j}| \leq \lambda_b\\
&rZ_{b,j} & &\text{ if }   |Z_{b,j}| >\lambda_b,
\end{aligned}\right.
\end{aligned}\right.
\end{equation}
in which $r$ is a constant.

After the coordinate descent steps, we can complete variable selection and get the sparse estimator $\hat{\boldsymbol{\beta}}_b$. Like the common penalty estimators, however, the online updating estimator also has a relatively large bias because of the use of penalty.
In order to get a bias-reduced estimator, after variable selection by the method proposed above, we  estimate the nonzero components again by renewable MLE in \cite{L2020b}. Thus, in the each step above, we use the corresponding bias-reduced penalty estimator as our choice of parameter estimation; such a bias-reduced estimator is still denoted by $\hat{\boldsymbol{\beta}}_b$.

In the above process,
only the diagonal elements of the negative Hessian matrix are considered. This is because maintaining a full Hessian matrix requires $O(p^2)$ memory space and $O(p^2)$ computational complexity, which is impractical for handling large-scale ultra-high dimensional data. We only use the diagonal elements of the negative Hessian matrix for reducing the computation and storage burdens.

\subsection{Tuning and Solution Paths}

The accuracy of selecting active variables is affected largely by the chosen value of the tuning parameter $\lambda_b$. Motivated by the BIC criterion provided by \cite{S1978}, in the online updating procedure of variable section and parameter estimation, we propose the following online updating BIC criterion at the $b$-th step to choose $\lambda_b$:
\begin{equation*}
\begin{aligned}
\lambda_b&=\arg\min_{\lambda \in\mathrm{R}} \\
&\left\{\hat{s}\ln(N_b)+\left((\hat{\boldsymbol{\beta}}_{b,\lambda}-\hat{\boldsymbol{\beta}}_{b-1})^{T}\left\{\sum_{t=1}^{b-1}\boldsymbol{J}_t(\hat{\boldsymbol{\beta}}_{t}; D_{t})\right\}(\hat{\boldsymbol{\beta}}_{b,\lambda}-\hat{\boldsymbol{\beta}}_{b-1})
-2\ell_{b}\left(\hat{\boldsymbol{\beta}}_{b,\lambda}; D_{b}\right)\right)\right\},
\end{aligned}
\end{equation*}
where $\hat{\boldsymbol{\beta}}_{b,\lambda}$ is an estimator given a $\lambda$ via online coordinate descent until the $b$-th batch.
The reason for this choice is given in Supplementary Material. However, it is difficult to realize the above optimization procedure because of the infinite select interval $\lambda \in\mathrm{R}$.
We see from \eqref{value_LASSO}, \eqref{value_SCAD} and \eqref{value_MCP} that $\hat{\beta}_{b,j}$ will stay zero if $|Z_{b,j}|< \lambda_b$, implying $\lambda_{max}=\max_{j}|Z_{b,j}|$. Thus, to easily realize the above optimization procedure, the tuning parameter is restricted in $(0,\lambda_{max})$. Consequently, a realizable BIC criterion for choosing tuning parameter is given by
\begin{equation}\label{bic}
\begin{aligned}
\lambda_b&=\arg\min_{\lambda \in (0, \lambda_{max})} \\
&\left\{\hat{s}\ln(N_b)+\left((\hat{\boldsymbol{\beta}}_{b,\lambda}-\hat{\boldsymbol{\beta}}_{b-1})^{T}\left\{\sum_{t=1}^{b-1}\boldsymbol{J}_t(\hat{\boldsymbol{\beta}}_{t}; D_{t})\right\}(\hat{\boldsymbol{\beta}}_{b,\lambda}-\hat{\boldsymbol{\beta}}_{b-1})-2\ell_{b}
\left(\hat{\boldsymbol{\beta}}_{b,\lambda}; D_{b}\right)\right)\right\}.
\end{aligned}
\end{equation} Obviously, it has the online updating form.
Amounts of simulations given in Section 4 will illustrate that the way of choosing $\lambda_b$ via \eqref{bic} is efficient and computationally simple.

\subsection{Summary of Algorithm}
Finally, we summarize the general algorithm and give the pseudocode for key numerical calculations in following Algorithm.
\begin{algorithm}[!h]
	\caption{Online Variable Selection and Parameter Estimation in GLM with Streaming Datasets}
	\begin{algorithmic}[1]
	\STATE Input: model $f(y|\boldsymbol{X}; \boldsymbol{\beta}_0)$, streaming data sets $D_1,\cdots, D_b, \cdots$
	\STATE Initialize $\hat{\boldsymbol{\beta}}_1$ in $D_{1}$ via offline LASSO/SCAD/MCP, and compute the components of $\boldsymbol{J}_1(\hat{\boldsymbol{\beta}}_1, D_1)$, where the corresponding indexes are in the selected active index set $\hat{S}_1$;
	\FOR{$b=2,3,\cdots$}
	\STATE  Choose $\hat{\lambda}_b$ via \eqref{bic};
	\STATE Compute $W_{b,j}=W_{b-1,j}+\boldsymbol{J}_b(\hat{\boldsymbol{\beta}}_{b-1}, D_b)_{[j,j]}$, $Z_{b,j}=\boldsymbol{U}_b\left(\hat{\boldsymbol{\beta}}_{b-1}; D_{b}\right)_{[j]}+\hat{\beta}_{b-1,j}W_{b,j}$, for $j \in \{1,\cdots,p\}$;
	\STATE Choose active index set via LASSO\eqref{value_LASSO}/SCAD\eqref{value_SCAD}/MCP\eqref{value_MCP}, $\hat{S}_b=\{j \in \{1,\cdots,p\}| \hat{\beta}_{b,j} \ne 0\}$;
	\STATE Estimate the parameters with indexes in $\hat{S}_b$ via renewable MLE in \citet{L2020b};
	\STATE  Compute the components of $\boldsymbol{J}_1(\hat{\boldsymbol{\beta}}_b, D_b)$, where the corresponding indexes are in the selected active index set $\hat{S}_b$; and update the cumulative negative Hessian matrix;
	\ENDFOR
	\STATE  Output $\hat{\boldsymbol{\beta}}_b$.
	\end{algorithmic}
\end{algorithm}
\section{Simulation Studies and Real Data Analysis}
In this section, we conduct simulation experiments to assess the performance of the proposed method in the settings of linear and logistic models. For a thorough evaluation, we compare our method with the following competitors:

(a) The total data offline LASSO proposed by \cite{T1996}, denoted by Total\_data\_LASSO. This method processes the entire data in a procedure of variable section and parameter estimation. Thus, it is not an online updating method. It is employed only as an object of reference;

(b) The total data offline SCAD proposed by \cite{F2001}, denoted by Total\_data\_SCAD. Like the Total\_data\_LASSO, this method is also employed as an object of reference;

(c) The total data offline MCP proposed by \cite{Z2010}, denoted by Total\_data\_MCP. Also this is an object of reference;

(d) The truncated SGD proposed by \cite{F2018}, denoted by Fan\_LASSO, Fan\_SCAD and Fan\_MCP respectively according to the methods in the burn-in step;

(e) Our online updating coordinate descent algorithms, denoted by Renew\_LASSO, Renew\_SCAD and Renew\_MCP respectively according to the types of the penalty function.

In the procedure of simulation, the parameter estimation is assessed by the average squared $\ell_2$ error, and the variable selection is evaluated by the following five criteria:

(a) the average number of the variables being selected, denoted by NV;

(b) the percentage of occasion on which all active variables are included in the selected model, denoted by IN;

(c) the percentage of occasion on which correct components are selected, denoted by CS;

(d) the type I error, irrelevant predictor inclusion rate, denoted by I;

(e) the type II error, active predictor exclusion rate, denoted by II.

\subsection{Example 1 (Gaussian linear model)}
\begin{figure}
\hypertarget{figure1}{}
  \centering
  \includegraphics[width=14cm]{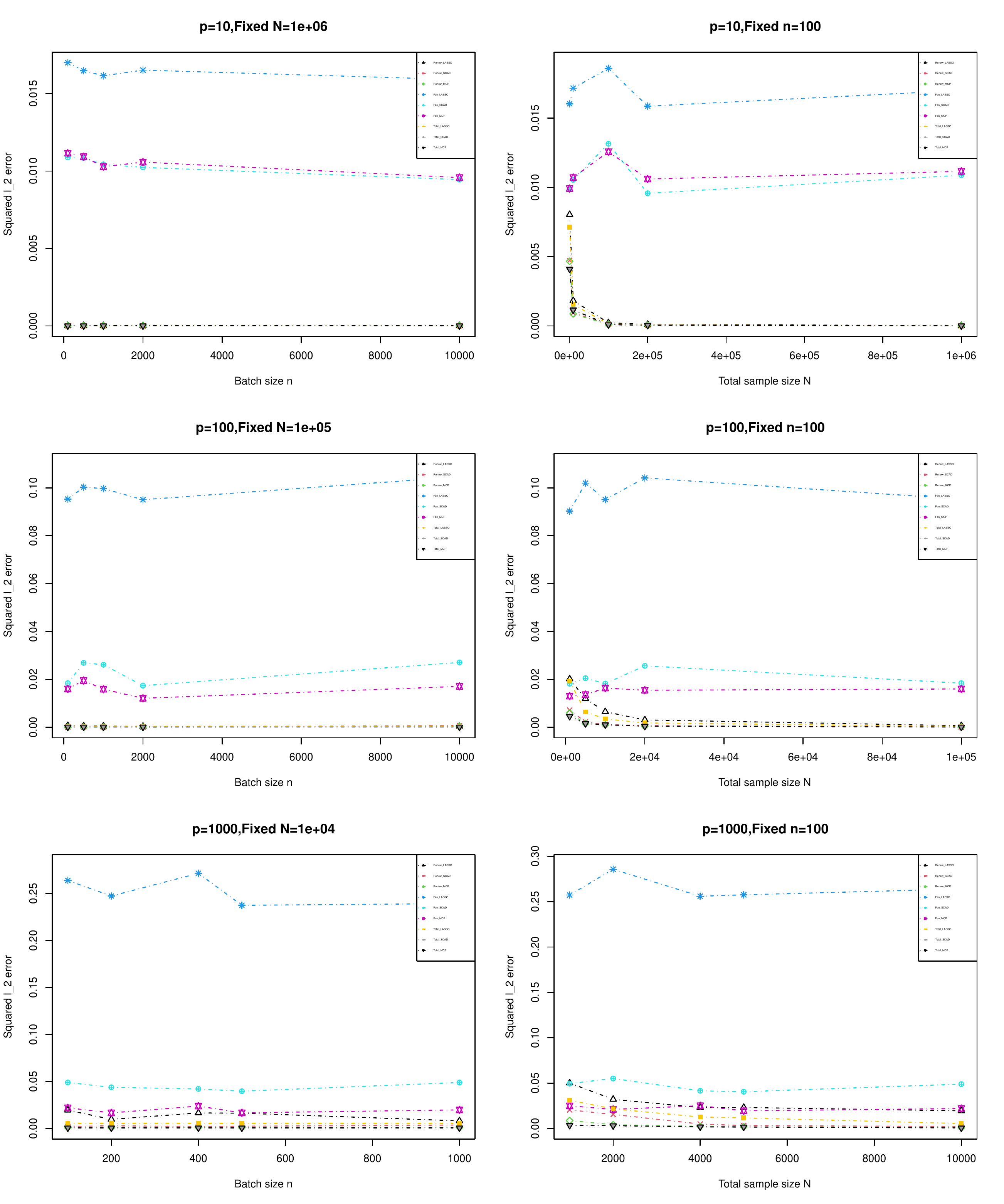}
  \caption{Parameter estimation results in the linear model.}
  \label{fig1}
\end{figure}

In this example, we generate the full data set $D_{B}^{*}$ and $N_{B}$ observations independently from the mean model $E(y_{ti}|\boldsymbol{x}_{ti})=g(\boldsymbol{x}_{ti}^{T}\boldsymbol{\beta}), i=1,\cdots, n_t; t=1,\cdots,B$, where  $g(\cdot)$ is the identical transformation, and $B$ is a terminal point.  For different dimensions of covariates, $p= 10, 100$ and $1000$, we always set the active index set as $S=\{1,2,3,4,5\}$, the true parameter vector as
$\boldsymbol{\beta}_{0}=(\underbrace{0.5,-0.5, 0.5,-0.5, 0.5}_{5}, \underbrace{0,0, \ldots, 0}_{p-5})^{\mathrm{T}}$, the intercept $\boldsymbol{x}_{ti,[1]} \equiv 1$, and $\boldsymbol{x}_{ti,[2:p]} \stackrel{iid}\sim \mathcal{N}(\boldsymbol{0}_{p-1},\boldsymbol{V}_{p-1})$, where $\boldsymbol{V}_{p-1}$ is a $(p-1) \times (p-1)$ compound symmetry covariance matrix with correlation $\rho=0.5$. As suggested in \cite{F2018}, for all renewable methods, we use the 1000 samples to give the starting values. In fact, it is always satisfied in the streaming data sets. For the methods of Fan\_LASSO, Fan\_SCAD and Fan\_MCP provided by  \cite{F2018}, we choose the stepwise as $\eta=\log(N_B)/N_B$. For SCAD and MCP, we choose the $r=3.7$ and $r=3$ as defaults respectively, which are suggested in \cite{F2001} and \cite{Z2010}. Assume that in every batches the numbers of observation data are the same as $n_t=n=N_B/B$. All the results are obtained via over 100 rounds of simulations.
We summarize the simulation results in Tables \ref{linearp10fixedN}-\ref{linearp1000fixedn}, and  Figure \ref{fig1}.

\begin{table}[!htbp]
\centering
\caption{Linear model p=10 Fixed N=$10^{6}$.}%
\label{linearp10fixedN}
\tiny{
\begin{tabular*}{400pt}{@{\extracolsep\fill}lllllll@{\extracolsep\fill}}
\toprule
Size&Method&NV&IN&CS&I&II  \\  
\toprule
{n=100} &Fan\_LASSO&7.85&1.00&0.07&0.285&0.000\\
{B=10000} &Renew\_LASSO&6.25&1.00&0.29&0.125&0.000\\
            &Fan\_SCAD &5.47&1.00&0.80&0.047&0.000\\
            &Renew\_SCAD &5.24&1.00&0.84&0.024&0.000\\
            &Fan\_MCP&5.51&1.00&0.77&0.051&0.000\\
            &Renew\_MCP&5.27&1.00&0.80&0.027&0.000\\
\midrule
{n=500} &Fan\_LASSO&7.79&1.00&0.06&0.279&0.000\\
{B=2000}&Renew\_LASSO&6.24&1.00&0.27&0.124&0.000\\
            &Fan\_SCAD&5.45&1.00&0.80&0.045&0.000\\
            &Renew\_SCAD &5.23&1.00&0.84&0.023&0.000\\
            &Fan\_MCP &5.42&1.00&0.79&0.042&0.000\\
            &Renew\_MCP&5.21&1.00&0.84&0.021&0.000\\
\midrule
{n=1000} &Fan\_LASSO&7.67&1.00&0.07&0.267&0.000\\
{B=1000}&Renew\_LASSO &6.00&1.00&0.30&0.100&0.000\\
            &Fan\_SCAD&5.48&1.00&0.79&0.048&0.000\\
            &Renew\_SCAD&5.22&1.00&0.83&0.022&0.000\\
            &Fan\_MCP&5.44&1.00&0.79&0.044&0.000\\
            &Renew\_MCP&5.21&1.00&0.84&0.021&0.000\\
\midrule
{n=2000} &Fan\_LASSO&7.96&1.00&0.03&0.296&0.000\\
{B=500}&Renew\_LASSO&6.08&1.00&0.34&0.108&0.000\\
            &Fan\_SCAD&5.46&1.00&0.81&0.046&0.000\\
            &Renew\_SCAD&5.17&1.00&0.88&0.017&0.000\\
            &Fan\_MCP&5.46&1.00&0.80&0.046&0.000\\
            &Renew\_MCP&5.14&1.00&0.89&0.014&0.000\\
\midrule
{n=10000}&Fan\_LASSO&7.79&1.00&0.06&0.279&0.000\\
{B=100}&Renew\_LASSO &5.91&1.00&0.36&0.091&0.000\\
            &Fan\_SCAD&5.35&1.00&0.85&0.035&0.000\\
            &Renew\_SCAD&5.01&1.00&0.99&0.001&0.000\\
            &Fan\_MCP&5.35&1.00&0.84&0.035&0.000\\
            &Renew\_MCP&5.02&1.00&0.98&0.002&0.000\\
\midrule
{N=$10^6$}&Total\_data\_LASSO&8.10&1.00&0.02&0.310&0.000\\
            &Total\_data\_SCAD&5.50&1.00&0.79&0.050&0.000\\
            &Total\_data\_MCP&5.60&1.00&0.72&0.060&0.000\\
\bottomrule
\end{tabular*}}
\end{table}

\begin{table}
\centering
\caption{Linear model p=10 Fixed n=100.}%
\label{linearp10fixedn}
\tiny{
\begin{tabular*}{400pt}{@{\extracolsep\fill}lllllll@{\extracolsep\fill}}
\toprule
Size&Method&NV&IN&CS&I&II  \\  
\toprule
{B=10} &Fan\_LASSO&7.94&1.00&0.03&0.294&0.000\\
{N=1000}&Renew\_LASSO  &6.21&1.00&0.29&0.121&0.000\\
            &Total\_data\_LASSO&8.04&1.00&0.03&0.304&0.000\\
            &Fan\_SCAD&5.38&1.00&0.83&0.038&0.000\\
            &Renew\_SCAD&5.28&1.00&0.84&0.028&0.000\\
             &Total\_data\_SCAD &5.36&1.00&0.83&0.036&0.000\\
            &Fan\_MCP &5.34&1.00&0.83&0.034&0.000\\
            &Renew\_MCP&5.25&1.00&0.85&0.025&0.000\\
             &Total\_data\_MCP &5.36&1.00&0.80&0.036&0.000\\
\midrule
{B=100} &Fan\_LASSO&7.83&1.00&0.05&0.283&0.000\\
{N=10000}&Renew\_LASSO &6.33&1.00&0.30&0.133&0.000\\
            &Total\_data\_LASSO&8.10&1.00&0.04&0.310&0.000\\
            &Fan\_SCAD &5.41&1.00&0.84&0.041&0.000\\
            &Renew\_SCAD &5.27&1.00&0.88&0.027&0.000\\
             &Total\_data\_SCAD &5.76&1.00&0.69&0.076&0.000\\
            &Fan\_MCP&5.36&1.00&0.83&0.036&0.000\\
            &Renew\_MCP&5.22&1.00&0.89&0.022&0.000\\
             &Total\_data\_MCP&5.67&1.00&0.68&0.067&0.000\\
\midrule
{B=1000} &Fan\_LASSO&7.96&1.00&0.05&0.296&0.000\\
{N=100000}&Renew\_LASSO  &6.39&1.00&0.19&0.139&0.000\\
            &Total\_data\_LASSO &8.01&1.00&0.07&0.301&0.000\\
            &Fan\_SCAD  &5.65&1.00&0.75&0.065&0.000\\
            &Renew\_SCAD  &5.31&1.00&0.78&0.031&0.000\\
             &Total\_data\_SCAD  &5.53&1.00&0.81&0.053&0.000\\
            &Fan\_MCP  &5.52&1.00&0.78&0.052&0.000\\
            &Renew\_MCP&5.29&1.00&0.80&0.029&0.000\\
             &Total\_data\_MCP &5.32&1.00&0.87&0.032&0.000\\
\midrule
{B=2000} &Fan\_LASSO &7.81&1.00&0.04&0.281&0.000\\
{N=200000}&Renew\_LASSO  &6.31&1.00&0.19&0.131&0.000\\
            &Total\_data\_LASSO &7.77&1.00&0.05&0.277&0.000\\
            &Fan\_SCAD &5.35&1.00&0.82&0.035&0.000\\
            &Renew\_SCAD  &5.24&1.00&0.84&0.024&0.000\\
             &Total\_data\_SCAD  &5.40&1.00&0.82&0.040&0.000\\
            &Fan\_MCP  &5.53&1.00&0.75&0.053&0.000\\
            &Renew\_MCP &5.25&1.00&0.80&0.025&0.000\\
             &Total\_data\_MCP  &5.27&1.00&0.84&0.027&0.000\\
\midrule
{B=10000} &Fan\_LASSO &7.85&1.00&0.07&0.285&0.000\\
{N=1000000}&Renew\_LASSO   &6.25&1.00&0.29&0.125&0.000\\
            &Total\_data\_LASSO &8.10&1.00&0.02&0.310&0.000\\
            &Fan\_SCAD  &5.47&1.00&0.80&0.047&0.000\\
            &Renew\_SCAD  &5.24&1.00&0.84&0.024&0.000\\
             &Total\_data\_SCAD  &5.50&1.00&0.79&0.050&0.000\\
            &Fan\_MCP  &5.51&1.00&0.77&0.051&0.000\\
            &Renew\_MCP &5.27&1.00&0.80&0.027&0.000\\
             &Total\_data\_MCP  &5.60&1.00&0.72&0.060&0.000\\
\bottomrule
\end{tabular*}}
\end{table}

\begin{table}
\centering
\caption{Linear model p=100 Fixed N=$10^{5}$.}%
\label{linearp100fixedN}
\tiny{
\begin{tabular*}{400pt}{@{\extracolsep\fill}lllllll@{\extracolsep\fill}}
\toprule
Size&Method&NV&IN&CS&I&II  \\   
\toprule
{n=100} &Fan\_LASSO&17.51&1.00&0.00&0.1251&0.0000\\
{B=1000} &Renew\_LASSO &7.22&1.00&0.25&0.0222&0.0000\\
            &Fan\_SCAD &5.88&1.00&0.69&0.0088&0.0000\\
            &Renew\_SCAD &5.43&1.00&0.73&0.0043&0.0000\\
            &Fan\_MCP&5.58&1.00&0.76&0.0058&0.0000\\
            &Renew\_MCP&5.30&1.00&0.78&0.0030&0.0000\\
\midrule
{n=500} &Fan\_LASSO&17.82&1.00&0.00&0.1282&0.0000\\
{B=200}&Renew\_LASSO&6.95&1.00&0.36&0.0195&0.0000\\
            &Fan\_SCAD&6.51&1.00&0.56&0.0151&0.0000\\
             &Renew\_SCAD &5.52&1.00&0.65&0.0052&0.0000\\
            &Fan\_MCP&5.73&1.00&0.76&0.0073&0.0000\\
            &Renew\_MCP&5.25&1.00&0.81&0.0025&0.0000\\
\midrule
{n=1000} &Fan\_LASSO&17.65&1.00&0.00&0.1265&0.0000\\
{B=100}&Renew\_LASSO &6.45&1.00&0.39&0.0145&0.0000\\
            &Fan\_SCAD&6.55&1.00&0.62&0.0155&0.0000\\
            &Renew\_SCAD&5.43&1.00&0.75&0.0043&0.0000\\
            &Fan\_MCP&5.43&1.00&0.81&0.0043&0.0000\\
            &Renew\_MCP&5.22&1.00&0.85&0.0022&0.0000\\
\midrule
{n=2000} &Fan\_LASSO&17.33&1.00&0.01&0.1233&0.0000\\
{B=50}&Renew\_LASSO&6.50&1.00&0.51&0.0150&0.0000\\
            &Fan\_SCAD&5.97&1.00&0.76&0.0097&0.0000\\
            &Renew\_SCAD&5.27&1.00&0.88&0.0027&0.0000\\
            &Fan\_MCP&5.34&1.00&0.85&0.0034&0.0000\\
            &Renew\_MCP&5.18&1.00&0.88&0.0018&0.0000\\
\midrule
{n=10000}&Fan\_LASSO&19.12&1.00&0.01&0.1412&0.0000\\
{B=10}&Renew\_LASSO &7.35&1.00&0.24&0.0235&0.0000\\
            &Fan\_SCAD&6.66&1.00&0.62&0.0166&0.0000\\
            &Renew\_SCAD&5.19&1.00&0.91&0.0019&0.0000\\
            &Fan\_MCP&5.59&1.00&0.74&0.0059&0.0000\\
            &Renew\_MCP&5.00&1.00&1.00&0.0000&0.0000\\
\midrule
{N=100000}&Total\_data\_LASSO&18.66&1.00&0.00&0.1366&0.0000\\
            &Total\_data\_SCAD&6.57&1.00&0.66&0.0157&0.0000\\
            &Total\_data\_MCP&5.60&1.00&0.78&0.0060&0.0000\\
\bottomrule
\end{tabular*}}
\end{table}

\begin{table}
\centering
\caption{Linear model p=100 Fixed n=100.}%
\label{linearp100fixedn}
\tiny{
\begin{tabular*}{400pt}{@{\extracolsep\fill}lllllll@{\extracolsep\fill}}
\toprule
Size&Method&NV&IN&CS&I&II  \\  
\toprule
{B=10} &Fan\_LASSO&16.98&1.00&0.00&0.1198&0.0000\\
{N=1000}&Renew\_LASSO  &7.51&1.00&0.29&0.0251&0.0000\\
            &Total\_data\_LASSO&17.64&1.00&0.00&0.1264&0.0000\\
            &Fan\_SCAD&5.86&1.00&0.76&0.0086&0.0000\\
            &Renew\_SCAD&5.56&1.00&0.78&0.0056&0.0000\\
             &Total\_data\_SCAD &6.10&1.00&0.72&0.0110&0.0000\\
            &Fan\_MCP&5.33&1.00&0.85&0.0033&0.0000\\
            &Renew\_MCP&5.30&1.00&0.86&0.0030&0.0000\\
             &Total\_data\_MCP&5.31&1.00&0.84&0.0031&0.0000\\
\midrule
{B=50} &Fan\_LASSO&18.36&1.00&0.00&0.1336&0.0000\\
{N=5000}&Renew\_LASSO &8.55&1.00&0.24&0.0355&0.0000\\
           &Total\_data\_LASSO&17.91&1.00&0.00&0.1291&0.0000\\
            &Fan\_SCAD&6.15&1.00&0.69&0.0115&0.0000\\
            &Renew\_SCAD &5.53&1.00&0.76&0.0053&0.0000\\
             &Total\_data\_SCAD &5.85&1.00&0.78&0.0085&0.0000\\
            &Fan\_MCP&5.41&1.00&0.82&0.0041&0.0000\\
            &Renew\_MCP&5.21&1.00&0.86&0.0021&0.0000\\
             &Total\_data\_MCP&5.33&1.00&0.87&0.0033&0.0000\\
\midrule
{B=100} &Fan\_LASSO&17.64&1.00&0.00&0.1264&0.0000\\
{N=10000}&Renew\_LASSO  &8.15&1.00&0.21&0.0315&0.0000\\
            &Total\_data\_LASSO &17.44&1.00&0.00&0.1244&0.0000\\
            &Fan\_SCAD  &5.87&1.00&0.77&0.0087&0.0000\\
            &Renew\_SCAD  &5.39&1.00&0.86&0.0039&0.0000\\
             &Total\_data\_SCAD  &6.61&1.00&0.67&0.0161&0.0000\\
            &Fan\_MCP  &5.57&1.00&0.77&0.0057&0.0000\\
            &Renew\_MCP&5.24&1.00&0.86&0.0024&0.0000\\
             &Total\_data\_MCP &5.60&1.00&0.84&0.0060&0.0000\\
\midrule
{B=200} &Fan\_LASSO &18.92&1.00&0.00&0.1392&0.0000\\
{N=20000}&Renew\_LASSO  &7.36&1.00&0.24&0.0236&0.0000\\
            &Total\_data\_LASSO &18.31&1.00&0.01&0.1331&0.0000\\
            &Fan\_SCAD &6.49&1.00&0.64&0.0149&0.0000\\
            &Renew\_SCAD  &5.54&1.00&0.74&0.0054&0.0000\\
             &Total\_data\_SCAD &6.18&1.00&0.72&0.0118&0.0000\\
            &Fan\_MCP  &5.45&1.00&0.79&0.0045&0.0000\\
            &Renew\_MCP  &5.18&1.00&0.87&0.0018&0.0000\\
             &Total\_data\_MCP  &5.33&1.00&0.85&0.0033&0.0000\\
\midrule
{B=1000} &Fan\_LASSO &17.51&1.00&0.00&0.1251&0.0000\\
{N=100000}&Renew\_LASSO   &7.22&1.00&0.25&0.0222&0.0000\\
            &Total\_data\_LASSO &18.66&1.00&0.00&0.1366&0.0000\\
            &Fan\_SCAD  &5.88&1.00&0.69&0.0088&0.0000\\
            &Renew\_SCAD  &5.43&1.00&0.73&0.0043&0.0000\\
             &Total\_data\_SCAD    &6.57&1.00&0.66&0.0157&0.0000\\
            &Fan\_MCP  &5.58&1.00&0.76&0.0058&0.0000\\
            &Renew\_MCP  &5.30&1.00&0.78&0.0030&0.0000\\
             &Total\_data\_MCP  &5.60&1.00&0.78&0.0060&0.0000\\
\bottomrule
\end{tabular*}}
\end{table}

\begin{table}
\centering
\caption{Linear model p=1000 Fixed N=$10^{4}$.}%
\label{linearp1000fixedN}
\tiny{
\begin{tabular*}{400pt}{@{\extracolsep\fill}lllllll@{\extracolsep\fill}}
\toprule
Size&Method&NV&IN&CS&I&II  \\  
\toprule
{n=100} &Fan\_LASSO&29.51&1.00&0.00&0.02451&0.00000\\
{B=100} &Renew\_LASSO &10.79&1.00&0.24&0.00579&0.00000\\
            &Fan\_SCAD &7.77&1.00&0.66&0.00277&0.00000\\
            &Renew\_SCAD &6.30&1.00&0.70&0.00130&0.00000\\
            &Fan\_MCP&5.68&1.00&0.79&0.00068&0.00000\\
            &Renew\_MCP&5.34&1.00&0.82&0.00034&0.00000\\
\midrule
{n=200} &Fan\_LASSO&27.67&1.00&0.00&0.02267&0.00000\\
{B=50}&Renew\_LASSO&9.35&1.00&0.21&0.00435&0.00000\\
            &Fan\_SCAD &7.48&1.00&0.67&0.00248&0.00000\\
            &Renew\_SCAD &6.24&1.00&0.71&0.00124&0.00000\\
            &Fan\_MCP&5.45&1.00&0.85&0.00045&0.00000\\
            &Renew\_MCP&5.25&1.00&0.86&0.00025&0.00000\\
\midrule
{n=400} &Fan\_LASSO&30.61&1.00&0.00&0.02561&0.00000\\
{B=25}&Renew\_LASSO &11.23&1.00&0.17&0.00623&0.00000\\
            &Fan\_SCAD &7.07&1.00&0.66&0.00207&0.00000\\
            &Renew\_SCAD &6.06&1.00&0.67&0.00106&0.00000\\
            &Fan\_MCP&5.74&1.00&0.76&0.00074&0.00000\\
            &Renew\_MCP&5.52&1.00&0.77&0.00052&0.00000\\
\midrule
{n=500} &Fan\_LASSO&26.06&1.00&0.01&0.02106&0.00000\\
{B=20}&Renew\_LASSO&11.46&1.00&0.13&0.00646&0.00000\\
           &Fan\_SCAD &7.35&1.00&0.73&0.00235&0.00000\\
            &Renew\_SCAD &5.84&1.00&0.74&0.00084&0.00000\\
            &Fan\_MCP&5.40&1.00&0.85&0.00040&0.00000\\
            &Renew\_MCP&5.31&1.00&0.86&0.00031&0.00000\\
\midrule
{n=1000}&Fan\_LASSO&27.01&1.00&0.00&0.02201&0.00000\\
{B=10}&Renew\_LASSO &8.74&1.00&0.12&0.00374&0.00000\\
            &Fan\_SCAD &7.89&1.00&0.69&0.00289&0.00000\\
            &Renew\_SCAD &5.97&1.00&0.78&0.00097&0.00000\\
            &Fan\_MCP&5.56&1.00&0.83&0.00056&0.00000\\
            &Renew\_MCP&5.19&1.00&0.91&0.00019&0.00000\\
           \midrule
{N=10000}&Total\_data\_LASSO&26.33&1.00&0.00&0.02133&0.00000\\
            &Total\_data\_SCAD&7.23&1.00&0.71&0.00223&0.00000\\
            &Total\_data\_MCP&5.58&1.00&0.76&0.00058&0.00000\\
\bottomrule
\end{tabular*}}
\end{table}

\begin{table}
\centering
\caption{Linear model p=1000 Fixed n=100.}%
\label{linearp1000fixedn}
\tiny{
\begin{tabular*}{400pt}{@{\extracolsep\fill}lllllll@{\extracolsep\fill}}
\toprule
Size&Method&NV&IN&CS&I&II  \\  
\toprule
{B=10} &Fan\_LASSO&28.54&1.00&0.00&0.02354&0.00000\\
{N=1000}&Renew\_LASSO  &10.27&1.00&0.21&0.00527&0.00000\\
            &Total\_data\_LASSO&27.33&1.00&0.00&0.02233&0.00000\\
            &Fan\_SCAD&7.73&1.00&0.58&0.00273&0.00000\\
            &Renew\_SCAD&7.39&1.00&0.58&0.00239&0.00000\\
             &Total\_data\_SCAD &6.71&1.00&0.69&0.00171&0.00000\\
            &Fan\_MCP&5.81&1.00&0.76&0.00081&0.00000\\
            &Renew\_MCP&5.68&1.00&0.77&0.00068&0.00000\\
             &Total\_data\_MCP &5.30&1.00&0.82&0.00030&0.00000\\
\midrule
{B=20} &Fan\_LASSO&31.12&1.00&0.00&0.02612&0.00000\\
{N=2000}&Renew\_LASSO &9.56&1.00&0.26&0.00456&0.00000\\
           &Total\_data\_LASSO&27.33&1.00&0.01&0.02233&0.00000\\
            &Fan\_SCAD&8.08&1.00&0.62&0.00308&0.00000\\
            &Renew\_SCAD&7.32&1.00&0.64&0.00232&0.00000\\
             &Total\_data\_SCAD &7.67&1.00&0.65&0.00267&0.00000\\
            &Fan\_MCP&5.55&1.00&0.74&0.00055&0.00000\\
            &Renew\_MCP&5.54&1.00&0.74&0.00054&0.00000\\
             &Total\_data\_MCP &5.60&1.00&0.77&0.00060&0.00000\\
             \midrule
{B=40} &Fan\_LASSO&28.16&1.00&0.01&0.02316&0.00000\\
{N=4000}&Renew\_LASSO &9.72&1.00&0.19&0.00472&0.00000\\
            &Total\_data\_LASSO&30.57&1.00&0.00&0.02557&0.00000\\
            &Fan\_SCAD&7.02&1.00&0.66&0.00202&0.00000\\
            &Renew\_SCAD&6.26&1.00&0.70&0.00126&0.00000\\
             &Total\_data\_SCAD &7.34&1.00&0.69&0.00234&0.00000\\
            &Fan\_MCP&5.74&1.00&0.72&0.00074&0.00000\\
            &Renew\_MCP&5.54&1.00&0.72&0.00054&0.00000\\
             &Total\_data\_MCP &5.54&1.00&0.81&0.00054&0.00000\\
\midrule
{B=50} &Fan\_LASSO&28.83&1.00&0.00&0.02383&0.00000\\
{N=5000}&Renew\_LASSO &10.86&1.00&0.26&0.00586&0.00000\\
            &Total\_data\_LASSO&28.89&1.00&0.00&0.02389&0.00000\\
            &Fan\_SCAD&7.21&1.00&0.61&0.00221&0.00000\\
            &Renew\_SCAD&5.93&1.00&0.67&0.00093&0.00000\\
             &Total\_data\_SCAD &6.63&1.00&0.67&0.00163&0.00000\\
            &Fan\_MCP&5.59&1.00&0.79&0.00059&0.00000\\
            &Renew\_MCP&5.39&1.00&0.82&0.00039&0.00000\\
             &Total\_data\_MCP &5.56&1.00&0.76&0.00056&0.00000\\
\midrule
{B=100} &Fan\_LASSO&29.51&1.00&0.00&0.02451&0.00000\\
{N=10000}&Renew\_LASSO  &10.76&1.00&0.24&0.00579&0.00000\\
            &Total\_data\_LASSO&26.33&1.00&0.00&0.02133&0.00000\\
            &Fan\_SCAD&7.77&1.00&0.66&0.00277&0.00000\\
            &Renew\_SCAD&6.30&1.00&0.70&0.00130&0.00000\\
             &Total\_data\_SCAD &7.23&1.00&0.71&0.00223&0.00000\\
            &Fan\_MCP&5.68&1.00&0.79&0.00068&0.00000\\
            &Renew\_MCP&5.34&1.00&0.82&0.00034&0.00000\\
             &Total\_data\_MCP &5.58&1.00&0.76&0.00058&0.00000\\
\bottomrule
\end{tabular*}}
\end{table}

We first report the variable selection results as follows.
From the Tables  \ref{linearp10fixedN},  \ref{linearp100fixedN},  \ref{linearp1000fixedN}, we have the following findings.

(1) When we fix the total sample size $N_B$ and vary the number of batches $B$, all the methods can choose the active variable sets such that $S \subseteq  \hat{S} $ for different dimensions of covariates, and our methods have smaller I, NV and larger CS. These illustrate that our methods could select all active variables and  less irrelevant variables than the other methods,  and our results are competitive with the total data offline methods.

(2) The  Fan\_LASSO, Fan\_SCAD and Fan\_MCP all have larger I, NV and smaller CS than ours.

From the Tables \ref{linearp10fixedn}, \ref{linearp100fixedn}, \ref{linearp1000fixedn}, we have the following findings.

(3) When we fix the sample size $n$ of every batches and vary the number of batch $B$, all the methods can select all the important variables.

(4) With larger total sample size $N_B$, our methods have the almost same selection results as the offline  Total\_data\_LASSO, Total\_data\_SCAD, Total\_data\_MCP. In this case, the  Fan\_LASSO, Fan\_SCAD and Fan\_MCP still include more irrelevant covariates.

For the parameter estimation, from the Figure \ref{fig1} we have the following observations.

(1) When we fix the total sample size $N_B$, and vary the numbers of batches and the dimensions of covariates, our methods have smaller average squared $\ell_2$ errors than the Fan\_LASSO, Fan\_SCAD and Fan\_MCP, and have almost same results as the offline  Total\_data\_LASSO, Total\_data\_SCAD, Total\_data\_MCP.

(2) When we fix the every batch sample size $n$, vary the dimensions of covariates, and increase numbers of batches,  our methods still have almost same results as the offline  Total\_data\_LASSO, Total\_data\_SCAD, Total\_data\_MCP. In this case, the Fan\_LASSO, Fan\_SCAD and Fan\_MCP have larger the average squared $\ell_2$ errors than ours and total data offline methods.

In summary, our methods obtain better variable selection and parameter estimation results than the Fan\_LASSO, Fan\_SCAD and Fan\_MCP, and almost the same results as the total data offline methods.

\subsection{Example 2 (Logistic model)}
In this example, the mean model is chosen as $E(y_{ti}|\boldsymbol{x}_{ti})=g(\boldsymbol{x}_{ti}^{T}\boldsymbol{\beta}), i=1,\cdots, n_t; t=1,\cdots,B$, where $g(\cdot)=\exp(\cdot)/\{1+\exp(\cdot)\}$. We choose $\boldsymbol{\beta}_{0}=(\underbrace{1,-1, 1,-1, 1}_{5}, \underbrace{0,0, \ldots, 0}_{p-5})^{\mathrm{T}}$ and set the others as the same as in Example 1. Note that the Fan\_LASSO, Fan\_SCAD and Fan\_MCP are only suitable for linear models. Thus, in the above logistic regression, we only compare our results with the offline LASSO, SCAD and MCP.

\begin{figure}
\hypertarget{figure2}{}
  \centering
  \includegraphics[width=14cm]{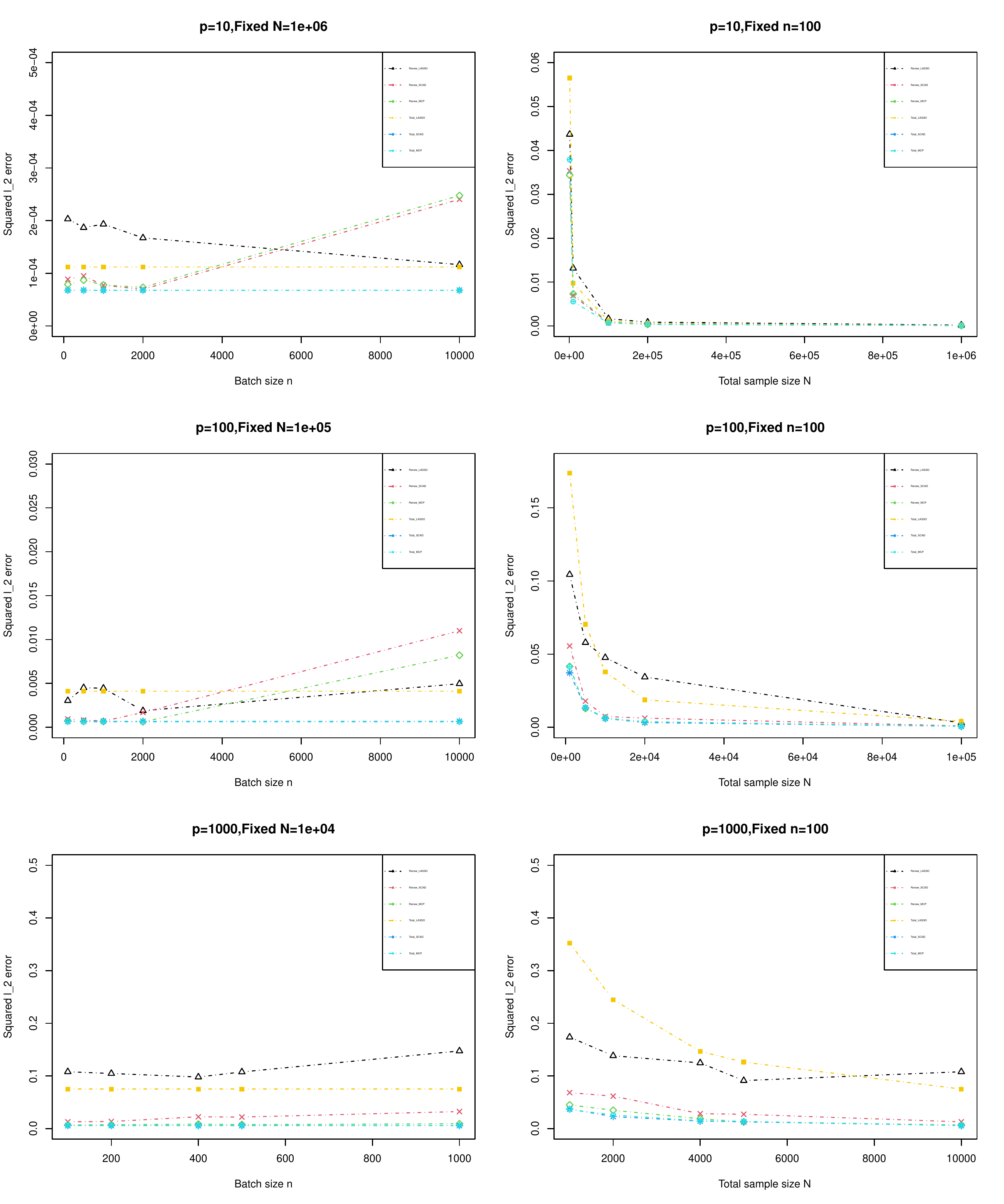}
  \caption{Parameter estimation results in the logistic model.}
  \label{fig2}
\end{figure}

We fix the total data size $N_B$  and  every batch sample size $n$ respectively. The selection results are summarized  in Tables \ref{logisticp10fixedN}-\ref{logisticp1000fixedn}, and the parameter estimation results are reported in Figure \ref{fig2}.

\begin{table}
\centering
\caption{Logistic model p=10 Fixed N=$10^{6}$.}%
\label{logisticp10fixedN}
\tiny{
\begin{tabular*}{400pt}{@{\extracolsep\fill}lllllll@{\extracolsep\fill}}
\toprule
Size&Method&NV&IN&CS&I&II  \\ 
\toprule
{n=100} &Renew\_LASSO &6.03&1.00&0.28&0.103&0.000\\
{B=10000} &Renew\_SCAD &5.25&1.00&0.81&0.025&0.000\\
            &Renew\_MCP&5.14&1.00&0.87&0.014&0.000\\
\midrule
{n=500} &Renew\_LASSO&5.89&1.00&0.36&0.089&0.000\\
{B=2000} &Renew\_SCAD&5.26&1.00&0.82&0.026&0.000\\
            &Renew\_MCP&5.22&1.00&0.82&0.022&0.000\\
\midrule
{n=1000} &Renew\_LASSO&5.97&1.00&0.33&0.097&0.000\\
{B=1000} &Renew\_SCAD &5.18&1.00&0.83&0.018&0.000\\
            &Renew\_MCP&5.20&1.00&0.84&0.020&0.000\\
\midrule
{n=2000} &Renew\_LASSO &5.77&1.00&0.38&0.077&0.000\\
{B=500} &Renew\_SCAD&5.02&1.00&0.98&0.002&0.000\\
            &Renew\_MCP&5.04&1.00&0.98&0.004&0.000\\
\midrule
{n=10000} &Renew\_LASSO &5.51&1.00&0.59&0.051&0.000\\
{B=100} &Renew\_SCAD&5.02&1.00&0.98&0.002&0.000\\
            &Renew\_MCP&5.03&1.00&0.97&0.003&0.000\\
\midrule
{N=1000000}&Total\_data\_LASSO&8.52&1.00&0.07&0.352&0.000\\
            &Total\_data\_SCAD&5.41&1.00&0.83&0.041&0.000\\
            &Total\_data\_MCP&5.30&1.00&0.86&0.030&0.000\\
\bottomrule
\end{tabular*}}
\end{table}

\begin{table}
\centering
\caption{Logistic model p=10 Fixed n=100.}%
\label{logisticp10fixedn}
\tiny{
\begin{tabular*}{400pt}{@{\extracolsep\fill}lllllll@{\extracolsep\fill}}
\toprule
Size&Method&NV&IN&CS&I&II  \\  
\toprule
{B=10} &Renew\_LASSO  &5.50&1.00&0.62&0.050&0.000\\
{N=1000} &Total\_data\_LASSO&8.37&1.00&0.02&0.337&0.000\\
             &Renew\_SCAD&5.37&1.00&0.80&0.037&0.000\\
             &Total\_data\_SCAD&5.43&1.00&0.84&0.043&0.000\\
            &Renew\_MCP&5.25&1.00&0.84&0.025&0.000\\
             &Total\_data\_MCP&5.38&1.00&0.86&0.038&0.000\\
\midrule
{B=100} &Renew\_LASSO &5.81&1.00&0.42&0.081&0.000\\
{N=10000} &Total\_data\_LASSO&8.26&1.00&0.01&0.326&0.000\\
            &Renew\_SCAD&5.13&1.00&0.90&0.013&0.000\\
             &Total\_data\_SCAD&5.45&1.00&0.77&0.045&0.000\\
            &Renew\_MCP&5.17&1.00&0.89&0.017&0.000\\
             &Total\_data\_MCP&5.20&1.00&0.86&0.020&0.000\\
\midrule
{B=1000} &Renew\_LASSO &6.00&1.00&0.31&0.100&0.000\\
{N=100000} &Total\_data\_LASSO&8.60&1.00&0.02&0.360&0.000\\
            &Renew\_SCAD&5.20&1.00&0.83&0.020&0.000\\
             &Total\_data\_SCAD&5.38&1.00&0.84&0.038&0.000\\
            &Renew\_MCP&5.31&1.00&0.80&0.031&0.000\\
             &Total\_data\_MCP&5.34&1.00&0.87&0.034&0.000\\
\midrule
{B=2000} &Renew\_LASSO&6.01&1.00&0.29&0.101&0.000\\
{N=200000} &Total\_data\_LASSO&8.39&1.00&0.02&0.339&0.000\\
             &Renew\_SCAD&5.23&1.00&0.82&0.023&0.000\\
             &Total\_data\_SCAD&5.59&1.00&0.76&0.059&0.000\\
            &Renew\_MCP&5.16&1.00&0.86&0.016&0.000\\
             &Total\_data\_MCP&5.54&1.00&0.74&0.054&0.000\\
\midrule
{B=10000} &Renew\_LASSO &6.03&1.00&0.28&0.103&0.000\\
{N=1000000} &Total\_data\_LASSO&8.52&1.00&0.07&0.352&0.000\\
            &Renew\_SCAD&5.25&1.00&0.81&0.025&0.000\\
           &Total\_data\_SCAD&5.41&1.00&0.83&0.041&0.000\\
            &Renew\_MCP&5.14&1.00&0.87&0.014&0.000\\
            &Total\_data\_MCP&5.30&1.00&0.86&0.030&0.000\\
\bottomrule
\end{tabular*}}
\end{table}

\begin{table}
\centering
\caption{Logistic model p=100 Fixed N=$10^{5}$.}%
\label{logisticp100fixedN}
\tiny{
\begin{tabular*}{400pt}{@{\extracolsep\fill}lllllll@{\extracolsep\fill}}
\toprule
Size&Method&NV&IN&CS&I&II  \\   
\toprule
{n=100} &Renew\_LASSO &5.82&1.00&0.58&0.0082&0.0000\\
{B=1000} &Renew\_SCAD  &5.33&1.00&0.76&0.0033&0.0000\\
            &Renew\_MCP &5.16&1.00&0.88&0.0016&0.0000\\
\midrule
{n=500} &Renew\_LASSO  &6.19&1.00&0.55&0.0119&0.0000\\
{B=200} &Renew\_SCAD &5.18&1.00&0.87&0.0018&0.0000\\
            &Renew\_MCP &5.11&1.00&0.92&0.0011&0.0000\\
\midrule
{n=1000} &Renew\_LASSO&6.21&1.00&0.57&0.0121&0.0000\\
{B=100} &Renew\_SCAD  &5.15&1.00&0.89&0.0015&0.0000\\
            &Renew\_MCP&5.15&1.00&0.92&0.0015&0.0000\\
\midrule
{n=2000} &Renew\_LASSO &5.59&1.00&0.68&0.0059&0.0000\\
{B=50} &Renew\_SCAD &5.30&1.00&0.84&0.0030&0.0000\\
            &Renew\_MCP &5.13&1.00&0.89&0.0013&0.0000\\
\midrule
{n=10000} &Renew\_LASSO &7.18&1.00&0.24&0.0218&0.0000\\
{B=10} &Renew\_SCAD&5.05&1.00&0.98&0.0005&0.0000\\
            &Renew\_MCP&5.00&1.00&1.00&0.0000&0.0000\\
\midrule
{N=100000}&Total\_data\_LASSO&21.56&1.00&0.00&0.1656&0.0000\\
            &Total\_data\_SCAD&5.98&1.00&0.74&0.0098&0.0000\\
            &Total\_data\_MCP&5.24&1.00&0.86&0.0024&0.0000\\
\bottomrule
\end{tabular*}}
\end{table}

\begin{table}
\centering
\caption{Logistic model p=100 Fixed n=100.}%
\label{logisticp100fixedn}
\tiny{
\begin{tabular*}{400pt}{@{\extracolsep\fill}lllllll@{\extracolsep\fill}}
\toprule
Size&Method&NV&IN&CS&I&II  \\   
\toprule
{B=10} &Renew\_LASSO  &6.05&1.00&0.56&0.0105&0.0000\\
{N=1000} &Total\_data\_LASSO&20.86&1.00&0.00&0.1586&0.0000\\
            &Renew\_SCAD&5.64&1.00&0.77&0.0064&0.0000\\
             &Total\_data\_SCAD &6.12&1.00&0.73&0.0112&0.0000\\
            &Renew\_MCP&5.27&1.00&0.83&0.0027&0.0000\\
             &Total\_data\_MCP &5.61&1.00&0.78&0.0061&0.0000\\
\midrule
{B=50} &Renew\_LASSO &6.40&1.00&0.58&0.0140&0.0000\\
{N=5000} &Total\_data\_LASSO&21.60&1.00&0.00&0.1660&0.0000\\
           &Renew\_SCAD&5.38&1.00&0.81&0.0038&0.0000\\
             &Total\_data\_SCAD &6.68&1.00&0.67&0.0168&0.0000\\
            &Renew\_MCP&5.17&1.00&0.89&0.0017&0.0000\\
             &Total\_data\_MCP &5.47&1.00&0.79&0.0047&0.0000\\
\midrule
{B=100} &Renew\_LASSO &6.73&1.00&0.55&0.0173&0.0000\\
{N=10000} &Total\_data\_LASSO&20.86&1.00&0.00&0.1586&0.0000\\
           &Renew\_SCAD&5.28&1.00&0.82&0.0028&0.0000\\
             &Total\_data\_SCAD &6.00&1.00&0.72&0.0100&0.0000\\
            &Renew\_MCP&5.14&1.00&0.90&0.0014&0.0000\\
             &Total\_data\_MCP &5.44&1.00&0.77&0.0044&0.0000\\
\midrule
{B=200} &Renew\_LASSO&6.74&1.00&0.45&0.0174&0.0000\\
{N=20000} &Total\_data\_LASSO &20.66&1.00&0.01&0.1566&0.0000\\
            &Renew\_SCAD&5.48&1.00&0.80&0.0048&0.0000\\
             &Total\_data\_SCAD &5.92&1.00&0.74&0.0092&0.0000\\
            &Renew\_MCP&5.20&1.00&0.87&0.0020&0.0000\\
             &Total\_data\_MCP &5.54&1.00&0.80&0.0054&0.0000\\
\midrule
{B=1000} &Renew\_LASSO &5.82&1.00&0.58&0.0082&0.0000\\
{N=100000} &Total\_data\_LASSO&21.56&1.00&0.00&0.1656&0.0000\\
            &Renew\_SCAD&5.33&1.00&0.76&0.0033&0.0000\\
             &Total\_data\_SCAD &5.98&1.00&0.74&0.0098&0.0000\\
            &Renew\_MCP&5.16&1.00&0.88&0.0016&0.0000\\
             &Total\_data\_MCP &5.24&1.00&0.86&0.0024&0.0000\\
\bottomrule
\end{tabular*}}
\end{table}

\begin{table}
\centering
\caption{Logistic model p=1000 Fixed N=$10^{4}$.}%
\label{logisticp1000fixedN}
\tiny{
\begin{tabular*}{400pt}{@{\extracolsep\fill}lllllll@{\extracolsep\fill}}
\toprule
Size&Method&NV&IN&CS&I&II  \\   
\toprule
{n=100} &Renew\_LASSO&7.88&1.00&0.53&0.00288&0.00000\\
{B=100} &Renew\_SCAD   &5.74&1.00&0.76&0.00074&0.00000\\
            &Renew\_MCP  &5.25&1.00&0.87&0.00025&0.00000\\
\midrule
{n=200} &Renew\_LASSO  &7.33&1.00&0.60&0.00233&0.00000\\
{B=50} &Renew\_SCAD &5.99&1.00&0.69&0.00099&0.00000\\
            &Renew\_MCP  &5.34&1.00&0.82&0.00034&0.00000\\
\midrule
{n=400} &Renew\_LASSO&7.41&1.00&0.37&0.00241&0.00000\\
{B=25} &Renew\_SCAD &6.38&1.00&0.55&0.00138&0.00000\\
            &Renew\_MCP&5.47&1.00&0.78&0.00047&0.00000\\
\midrule
{n=500} &Renew\_LASSO &8.33&1.00&0.33&0.00333&0.00000\\
{B=20} &Renew\_SCAD &6.40&1.00&0.63&0.00140&0.00000\\
            &Renew\_MCP &5.34&1.00&0.84&0.00034&0.00000\\
\midrule
{n=1000} &Renew\_LASSO &7.45&1.00&0.23&0.00245&0.00000\\
{B=10} &Renew\_SCAD&6.04&1.00&0.69&0.00104&0.00000\\
            &Renew\_MCP&5.26&1.00&0.89&0.00026&0.00000\\
\midrule
{N=10000}&Total\_data\_LASSO&36.46&1.00&0.00&0.03146&0.00000\\
            &Total\_data\_SCAD&7.10&1.00&0.71&0.00210&0.00000\\
            &Total\_data\_MCP&5.45&1.00&0.81&0.00045&0.00000\\
\bottomrule
\end{tabular*}}
\end{table}

\begin{table}
\centering
\caption{Logistic model p=1000 Fixed n=100.}%
\label{logisticp1000fixedn}
\tiny{
\begin{tabular*}{400pt}{@{\extracolsep\fill}lllllll@{\extracolsep\fill}}
\toprule
Size&Method&NV&IN&CS&I&II  \\ 
\toprule
{B=10} &Renew\_LASSO  &5.58&1.00&0.71&0.00058&0.00000\\
{N=1000} &Total\_data\_LASSO&34.95&1.00&0.00&0.02995&0.00000\\
            &Renew\_SCAD&5.75&1.00&0.66&0.00075&0.00000\\
             &Total\_data\_SCAD &7.78&1.00&0.62&0.00278&0.00000\\
            &Renew\_MCP &5.44&1.00&0.76&0.00044&0.00000\\
             &Total\_data\_MCP &5.55&1.00&0.80&0.00055&0.00000\\
\midrule
{B=20} &Renew\_LASSO  &6.16&1.00&0.61&0.00116&0.00000\\
{N=2000} &Total\_data\_LASSO&32.52&1.00&0.00&0.02752&0.00000\\
           &Renew\_SCAD&6.16&1.00&0.59&0.00116&0.00000\\
             &Total\_data\_SCAD &7.76&1.00&0.59&0.00276&0.00000\\
            &Renew\_MCP &5.49&1.00&0.78&0.00049&0.00000\\
             &Total\_data\_MCP &5.75&1.00&0.74&0.00075&0.00000\\
\midrule
{B=40} &Renew\_LASSO  &6.34&1.00&0.58&0.00134&0.00000\\
{N=4000} &Total\_data\_LASSO&39.35&1.00&0.00&0.03435&0.00000\\
           &Renew\_SCAD&5.90&1.00&0.68&0.00090&0.00000\\
             &Total\_data\_SCAD &8.20&1.00&0.61&0.00320&0.00000\\
            &Renew\_MCP &5.49&1.00&0.73&0.00049&0.00000\\
             &Total\_data\_MCP &5.69&1.00&0.79&0.00069&0.00000\\
\midrule
{B=50} &Renew\_LASSO &6.66&1.00&0.62&0.00166&0.00000\\
{N=5000} &Total\_data\_LASSO &36.27&1.00&0.00&0.03127&0.00000\\
           &Renew\_SCAD&6.06&1.00&0.67&0.00106&0.00000\\
             &Total\_data\_SCAD &8.04&1.00&0.64&0.00304&0.00000\\
            &Renew\_MCP &5.37&1.00&0.78&0.00037&0.00000\\
             &Total\_data\_MCP &5.75&1.00&0.77&0.00075&0.00000\\
\midrule
{B=100} &Renew\_LASSO  &7.88&1.00&0.53&0.00288&0.00000\\
{N=10000} &Total\_data\_LASSO&36.46&1.00&0.00&0.03146&0.00000\\
            &Renew\_SCAD&5.74&1.00&0.76&0.00074&0.00000\\
             &Total\_data\_SCAD&7.10&1.00&0.71&0.00210&0.00000\\
            &Renew\_MCP &5.25&1.00&0.87&0.00025&0.00000\\
             &Total\_data\_MCP&5.45&1.00&0.81&0.00045&0.00000\\
\bottomrule
\end{tabular*}}
\end{table}

We report the variable selection results firstly.  From the Tables  \ref{logisticp10fixedN},  \ref{logisticp100fixedN}, \ref{logisticp1000fixedN}, we have the following findings.

(1) All methods could achieve  $S \subseteq  \hat{S}$ when we fix the total sample size $N_B$ and vary the number of batches $B$. Our Renew\_SCAD and Renew\_MCP almost have the same NV, CS and I  as those of the offline  Total\_data\_SCAD and Total\_data\_MCP.

(2) The method Renew\_LASSO has smaller NV and I and larger CS than the offline Total\_data\_LASSO.

From the Tables  \ref{logisticp10fixedn},  \ref{logisticp100fixedn}, \ref{logisticp1000fixedn}, we have the following observations.

(3) When we fix the every batch sample sizes $n$, vary the numbers of batches and dimensions of covariates, all the methods can select all the active variables.

(4) It is seen that the  Renew\_LASSO  could select less irrelevant variables than the offline Total\_data\_LASSO.

For parameter estimation, from the Figure \ref{fig2}, we have the following observations.

(1) When we fix the total sample size $N_B$, vary the dimensions of covariates and increase the numbers of batches, the results of our methods are almost accordance with the results of offline Total\_data\_LASSO, Total\_data\_SCAD and Total\_data\_MCP.

(2) When we fix the every batch sample size $n$, vary dimensions of covariates, and increase numbers of batches, our methods still almost have the same performances as the offline methods.

In summary, our methods almost have the same behaviors  as the total data offline methods. Especially, the  Renew\_LASSO  performances are better than the offline  Total\_data\_LASSO.  In the following subsection of real data analysis, we will further illustrate the properties  via the corresponding numerical results.

\subsection{Real Data Analysis}
In the real data analysis, we apply our renewable variable selection methods, such as renewable LASSO, renewable SCAD, renewable MCP, to the following real dataset. It is the YearPredictionMSD dataset from the UCI Machine Learning Repository \cite{D2019}. This dataset, with 90 features, 463715 training samples and 51630 testing samples, is about the prediction of the release year of a song from audio features. For all online updating methods, we  give the starting values by $m=3715$ training samples. We use the $R^2$ values of estimators from various methods to confirm the advantages of the new methods, where $R^2=1-\sum_{j=1}^{n_{test}}(y_j-\hat{y}_j)^2/\sum_{i=1}^{n_{test}}(y_j-\overline{y}_j)^2$, $n_{test}$ is the total sample size in testing dataset, $\hat{y}_j$ is the predicted value of $y_j$, and $\overline{y}_j$ is the sample mean of $y_j$'s. We summarize the $R^2$ values  in Table \ref{realdata}, in which the Total\_data\_LS is the least square estimator using the total data.

\begin{table}
\centering
\caption{Results for YearPredictionMSD data set.}%
\label{realdata}
\tiny{
\begin{tabular*}{400pt}{@{\extracolsep\fill}llllll@{\extracolsep\fill}}
\toprule
Size&Method&$R^2$&Size&Method&$R^2$ \\   
\toprule
{n=230} &Fan\_LASSO&0.17104&{n=4600} &Fan\_LASSO&0.17104\\
{B=2000} &Renew\_LASSO&0.23167&{B=100} &Renew\_LASSO&0.23170\\
            &Fan\_SCAD &0.17117&&Fan\_SCAD &0.17061\\
             &Renew\_SCAD &0.23169&&Renew\_SCAD&0.23130\\
            &Fan\_MCP&0.17210&&Fan\_MCP&0.17210\\
            &Renew\_MCP&0.23180&&Renew\_MCP&0.23142\\
\midrule
{n=460} &Fan\_LASSO&0.17104&{n=9200} &Fan\_LASSO&0.17224\\
{B=1000} &Renew\_LASSO&0.23168&{B=50} &Renew\_LASSO&0.23175\\
             &Fan\_SCAD &0.17117&&Fan\_SCAD &0.17117\\
             &Renew\_SCAD &0.231723&&Renew\_SCAD&0.22920\\
            &Fan\_MCP&0.17147&&Fan\_MCP&0.17210\\
            &Renew\_MCP&0.23169&&Renew\_MCP&0.23013\\
\midrule
{n=920} &Fan\_LASSO&0.17224&{N=$463715$}&Total\_data\_LS&0.23169\\
{B=500} &Renew\_LASSO&0.23166&&&\\
            &Fan\_SCAD &0.17061&&&\\
             &Renew\_SCAD &0.23163&&&\\
            &Fan\_MCP&0.17210&&&\\
             &Renew\_MCP&0.23167&&&\\
\bottomrule
\end{tabular*}}
\end{table}

The results of the table clearly show that the $R^2$ values of the our renewable selection and estimators are larger than the other online updating estimators in general, and ours almost have the same results as the dotal data least squared estimators. Thus, our methods are the best choices for fitting the real data.

\section{Conclusion}

In the previous sections of this paper, for high-dimensional GLMs, we proposed a novel online updating objective function and related algorithm to achieve variable selection and parameter estimation. This is a unified framework because it can be applied to various penalty likelihoods with differentiable or non-differentiable penalty functions. Unlike the truncated method, ours choose the tuning parameter via data-driven criteria, such as BIC, and the proposed BIC criterion has an online updating form. Moreover, the proposed strategy can achieve the estimation and selection consistencies, and the oracle property, and these theoretical properties always holds, without any constraint on the number of the data batches.
Through various simulation studies, it is demonstrated that our proposed methods can recover the support of the true features, efficiently and accurately, and are much better the competitors and are competitive with the objects of reference by total data methods.

However, still there are some important issues needed to be investigated in the future. For example, in the environments of streaming data, usually the distributions of data will change or have  heterogeneity in different data periods.
How to design the objective function and the algorithm to adapt to the situation is a very significant issue in the streaming data. Another example is how to extend our methods into nonlinear models and nonparametric models. These are interesting issues and are worth further study in the future.
\bigskip
\begin{center}
{\large\bf SUPPLEMENTARY MATERIAL}
\end{center}

\begin{description}

\item[Supplementary Material:] Supplementary Material includes the details of online updating BIC criterion and some proofs of a lemma and theorems. (.pdf file)

\end{description}

\begin{spacing}{1.0}
\bibliographystyle{agsm}
\bibliography{online}
\end{spacing}
\end{document}